\documentclass[useAMS,usenatbib]{mn2e}
\usepackage{txfonts}
\usepackage{natbib}
\usepackage{graphicx}
\usepackage{times}

\usepackage{xspace}
\usepackage{color,url}
\usepackage{a_macros}

\bibpunct{(}{)}{;}{a}{}{,}
\newcommand{\au}{\hbox{AU~Mon }}
\newcommand{\aue}{\hbox{AU~Mon}}

\newcommand{\ubv}{\hbox{$U\!B{}V$}}
\newcommand{\bv}{\hbox{$B\!-\!V$}}

\newcommand{\uvby}{\hbox{$uvby$}}
\newcommand{\hp}{\hbox{$H_{\rm p}$}}
\newcommand{\oc}{\hbox{$O\!-\!C$}}

\newcommand{\p}{$\pm$}

\newcommand{\m}{$^{\rm m}\!\!.$}
\newcommand{\ANG}{\accent'27A}

\newcommand{\kms}{km\,s$^{-1}$ }
\newcommand{\ks}{km\,s$^{-1}$}
\newcommand{\vsin}{$v$~sin~$i$ }

\newcommand{\tef}{$T_{\rm eff}$ }
\newcommand{\lgg}{{\rm log}~$g$ }
\newcommand{\logg}{{\rm log}~$g$}

\newcommand{\ha}{H$\alpha$ }

\newcommand{\phoebe}{{\tt PHOEBE} }
\newcommand{\phoe}{{\tt PHOEBE}}
\newcommand{\korel}{{\sc korel}\xspace}
\newcommand{\corot}{{CoRoT}\xspace}
\newcommand{\teff}{$T_{\rm eff}$}

\newcommand{\vsini}{$v$~sin~$i$}
\def\cd{\ensuremath{{\rm d}^{-1}}\xspace}

\def\mHz{\ensuremath{\mu{\rm Hz}}\xspace}

\DeclareRobustCommand{\ion}[2]{%
\relax\ifmmode
\ifx\testbx\f@series
{\mathbf{#1\,\mathsc{#2}}}\else
{\mathrm{#1\,\mathsc{#2}}}\fi
\else\textup{#1\,{\mdseries\textsc{#2}}}%
\fi}

\begin{document}

\title[CoRoT photometry and high-resolution spectroscopy of the interacting eclipsing binary AU Mon]
{CoRoT photometry and high-resolution spectroscopy of the interacting
eclipsing binary AU~Mon\thanks{Based on photometry collected by the CoRoT
space mission as well as
spectroscopy obtained with the FEROS spectrograph attached to the 2.2-m
telescope at ESO, La Silla, Chile, under the ESO Large Programme LP178.D-0361,
 and with the SOPHIE spectrograph of the
Observatoire de Haute-Provence (France). The CoRoT space mission was developed and is
operated by the French space agency CNES, with participation of ESA's RSSD and
Science Programmes, Austria, Belgium, Brazil, Germany, and Spain.  
Based on observations collected at the Centro Astron\'omico Hispano Alem\'an
(CAHA) at Calar Alto, operated jointly by the Max-Planck Institut f\"ur
Astronomie and the Instituto de Astrof\'{\i}sica de Andaluc\'ia (CSIC).
}
}
\author[M. Desmet et al.]{M.~Desmet$^{1}$\thanks{E-mail: maarten.desmet@ster.kuleuven.be},
Y.~Fr\'emat$^{2}$,
F.~Baudin$^{3}$,
P.~Harmanec$^{4}$,
P.~Lampens$^{2}$,
E.~Janot Pacheco$^{5}$,
\newauthor
M.~Briquet$^{1}$,
P.~Degroote$^{1}$,
C.~Neiner$^{6}$,
P.~Mathias$^{7}$,
E.~Poretti$^{8}$,
M.~Rainer$^{8}$,
\newauthor
K.~Uytterhoeven$^{8,9}$,
P.~J.~Amado$^{10}$,
J.-C.~Valtier$^{7}$,
A.~{Pr{\v s}a}$^{11,12}$,
C.~Maceroni$^{13}$
\newauthor
and C.~Aerts$^{1,14}$
\\
$^1$ Instituut voor Sterrenkunde, K.U.Leuven,
Celestijnenlaan~200~D, B-3001 Leuven, Belgium\\
$^2$ Royal Observatory of Belgium, 3 Avenue circulaire, B-1180 Brussels,
Belgium\\
$^3$ Institut d'Astrophysique Spatiale, CNRS/Universit\'e Paris XI UMR 8617, 91405
Orsay, France\\
$^4$ Astronomical Institute of the Charles University, Faculty of Mathematics and
Physics, V~Hole\v{s}ovi\v{c}k\'ach 2, CZ-180 00 Praha 8, Czech Republic\\
$^5$ Universidade de S\~ao Paulo, Instituto de Astronomia, Geof\'isica e Ci\^encias
Atmosf\'ericas - IAG, Departamento de Astronomia,
Rua do Mat\~ao, \\
1226 - 05508-900 S\~ao Paulo, Brazil\\
$^6$ Observatoire de Paris - Section de Meudon,
Place Jules Janssen 5,
92195 Meudon cedex, France\\
$^7$ UNS, CNRS, OCA, Campus Valrose, UMR 6525 H. Fizeau, F-06108 Nice Cedex
2, France\\
$^8$ INAF - Osservatorio Astronomico di Brera, via Bianchi 46, 23807 Merate (LC),
Italy\\
$^9$ Laboratoire AIM, CEA/DSM-CNRS-Universit\'e Paris Diderot; CEA, IRFU, SAp,
centre de Saclay, F-91191, Gif-sur-Yvette, France\\
$^{10}$ Instituto de Astrof\'{\i}sica de Andaluc\'{\i}a (CSIC), Granada, Spain\\
$^{11}$Villanova University, Dept. Astron. Astrophys., 800 E Lancaster Ave,
Villanova, PA 19085, USA\\
$^{12}$ University of Ljubljana, Dept. of Physics, Jadranska
19, SI-1000 Ljubljana, Slovenia\\
$^{13}$ Osservatorio Astronomico di Roma, via Frascati 33, I-00040 Monteporzio
(RM), Italy\\
$^{14}$ Department of Astrophysics, IMAPP,
Radboud University Nijmegen, PO Box 9010, 6500 GL
Nijmegen, the Netherlands
}

\date{Received \today; Accepted...}

\pagerange{\pageref{firstpage}--\pageref{lastpage}} \pubyear{2009}

\maketitle

\label{firstpage}

\begin{abstract}
Analyses of very accurate CoRoT space photometry, past Johnson $V$
photoelectric photometry and high-resolution \'echelle spectra led to the
determination of improved and consistent fundamental stellar properties of
both components of AU~Mon.  We derived new, accurate ephemerides for both the
orbital motion (with a period of 11\fd113) and the long-term, overall
brightness variation (with a period of 416\fd9) of this strongly interacting
Be + G semi-detached binary. It is shown that this long-term variation must be
due to attenuation of the total light by some variable circumbinary material.
We derived the binary mass ratio $M_{\rm G}/M_{\rm B}$ = 0.17\p0.03  based on
the assumption that the G-type secondary fills its Roche lobe and rotates
synchronously. Using this value of the  mass ratio as
well as the radial velocities of the G-star, we obtained a consistent
light curve model and improved estimates of the stellar masses, radii,
luminosities and effective temperatures. 
We demonstrate that the observed lines of
the B-type  primary may not be of photospheric origin.
We also  discover rapid and periodic
light changes visible in the high-quality residual CoRoT light curves. 
AU~Mon is put into perspective by a comparison with known
binaries exhibiting long-term cyclic light changes. 
\end{abstract}



\section{Introduction}
AU~Monocerotis was selected as one of the  few known binary targets in  the
asteroseismology field of the French-European \corot space mission
\citep[Convection, Rotation and planetary Transits,][see ``the \corot
book'']{2006ESASP1306.....F} during the Initial Run (IRa01).  We present here a detailed
study of \au based on a  long, uninterrupted series of high-precision \corot
photometry obtained in 2008 as well as on high-dispersion \'echelle spectra
secured at three ground-based observatories in 2007.

AU~Mon (HD~50846, HIP~33237) is an interacting, eclipsing and double-lined
spectroscopic binary consisting of a Be star and an evolved late-type giant
star which in all probability fills its Roche lobe and loses matter towards
the Be star. To avoid confusion, we shall hereafter denote the mass-gaining
and the mass-losing components of \au  as the B- and G-stars respectively (We
will indeed demonstrate that the cool component is a G-type star rather than
an F-type star which was its classification based on photographic spectra).
The orbit is circular and the orbital period is 11\fd11302 \citep[see O-C
diagrams\footnote{\url{http://www.as.wsp.krakow.pl/o-c/index.php3}}\,of]
[]{krei2004}.
\au is a rare Algol-type system member of the W~Serpentis subclass,
phenomenologically defined by \citet{Pla78} and \citet{Pla80}. These binaries
are sometimes also called ``massive or hot Algols''. According to current
knowledge, they are characterized by a semi-detached configuration and several
distinct components of circumstellar matter: a ``cloud'' of very hot plasma
manifesting itself by emission lines in far-UV spectra and probably
located outside the orbital plane, a disk, possibly optically thick, 
which mimics a
false photosphere with a lower \teff\ than the central star, a gas stream
between both components, and a ``hot-line region'' arising from the
interaction of the gas stream and the disk/cloud complex \citep{bis2007}.  We
mention only some published studies of W~Ser binaries, relevant to our present
work. \citet{eli90} studied centimeter observations of 6 different W~Ser
binaries and found very strong evidence for circumstellar matter around these
systems. The same was claimed for SX~Cas \citep{ander89} and RX~Cas
\citep{ander89} on the basis of photometry and spectroscopy.
A system which is similar to \au is W~Cru, consisting of a B-star and a G
supergiant. \citet{pavl06} found this object to have 
a very extended accretion disk
with a clumpy rim, and suggested that the clumpiness may account for the
light curve distortions and asymmetries, as well as for secular changes.


\section{Previous knowledge about AU~Mon}
\subsection{Light changes, orbital period and ephemeris}
The eclipsing nature of \au was discovered by \citet{hoffm31}.  The first
ephemeris was published by \citet{florja37}. \citet{lause38} derived a
slightly longer value of the period.
\citet{lorenzi77} observed it and concluded that  
its light curve was undergoing rapid cyclic changes.
He suggested that this could be due to a fast apsidal motion with a period of
243\fd2 but also pointed out problems with such an interpretation. Shortly
thereafter, \citet{cerruti77} came up with the correct explanation: the
brightness of the whole system varied secularly. To understand the phenomenon,
\citet{lorenzi80a} secured 2616 $V$ differential observations between 1976 and
1979 at two observatories. From a detailed analysis of these data,
\citet{lorenzi80b} determined that the brightness of \au varied
cyclically with a period of 411 days and a peak-to-peak amplitude of about
0\m2. Taking this variation into account, he derived an improved ephemeris for
the binary system: 
\begin{equation} T_{\rm Min\,I}={\rm JD}\,2442801.3752(51)
+ 11\fd1130371(68) \times E,\label{efe1} 
\end{equation} 
\noindent which has
since been used by various other investigators.  \citet{lorenzi85} analyzed
new observations from 1983 and 1984 and confirmed the periodic character of
the brightness variations of \aue. A regularly updated ephemeris, based on all
compiled times of minima, can be found in the database of the \oc\ diagrams
\citep{krei2004}\footnotemark[\value{footnote}].

\subsection{Spectroscopic investigations }\label{sec:spec}
The first spectroscopic study aimed at the determination of the orbital
elements was carried out by \citet{sahade45}. Helped in the spectral
classification by Dr. Morgan, they concluded from the spectra, at maximum
light and during the eclipse, that the primary is a B5 main-sequence star and
the secondary has a spectral class near F0. In a later study, \citet{sahade82}
investigated not only ground-based but also the far-UV spectra of the binary
obtained with the International Ultraviolet Explorer (hereafter IUE). Faint
lines of the secondary were detected in the \ion{Na}{i} 5889 \& 5895\,\ANG\
doublet \citep{popper62} and later on measured quantitatively
\citep{popper89}. \citet{popper62} also noted the presence of a double \ha
emission. The first study, partly based on electronic spectra, was published
by \citet{sahade97}.  A previous determinations of the orbital elements of \au
is listed in Table~\ref{tab:sah}. 

\begin{table}
\vspace{0.1cm}
\caption{Spectroscopic orbital elements for AU~Mon derived by  
\citet{sahade97} in a spectroscopic analysis.}\label{tab:sah}
\tabcolsep=3pt
\begin{tabular}{lc}
\hline\hline\noalign{\smallskip}
Element & Value \\
\noalign{\smallskip}\hline\noalign{\smallskip}
$a_B\sin i\,(10^6 \rm km)$ & $6.6\pm0.39$\\
$a_G\sin i\,(10^6 \rm km)$ & $2.2 \pm 0.41$\\
$K_B$ (\kms) &  $43\pm2.5$\\
$K_G$ (\kms) &  $147\pm3.0$\\
$e$          & $0.06\pm0.02$\\
$\omega\,(^\circ)$ &$204\pm22$  \\
$\gamma$ (\kms) &  $2.1\pm1.5$\\
$M_B\,(M_\odot)$ & $6.1\pm0.61$\\
$M_G\,(M_\odot)$ & $1.8\pm0.53$\\
\hline
\end{tabular}
\end{table}

The far-UV spectra of \au were obtained by IUE and studied by
\citet{pol82, pet82, pet88,1989Ap.....30..407E, pet91, pet94a, pet94b, pet96} and
\citet{pet98}. \citet{pet82} confirmed that the gas stream between
the components is seen in the UV resonance lines of \ion{N}{v} and
\ion{C}{iv} in the orbital phase range 0.85 -- 0.93, where phase zero
corresponds to the time of primary eclipse. 
\citet{1989Ap.....30..407E} found the electron density of the gas in the
envelope to be $n_e\sim2\cdot10^{11}\,{\rm cm}^{-3}$, and the electron
temperature to be ${\rm T}_e\sim20\,000\,{\rm K}$.
\citet{pet88} found a correlation
between the instantaneous mass transfer rate and the line and
continuum spectrum along the 411-d cycle. At the maximum of the
mass transfer rate, the object was fainter and the accretion disk
denser while, at the brightness maximum, the high-temperature
plasma was more prominent in the UV spectral lines. Analyzing the
spectral energy distribution (SED) from the low-dispersion IUE
spectra, \cite{pet91} concluded that the optical brightness
variations of the system are due to a 1200~K variation in the
photospheric effective temperature of the Be primary. She also
suggested the presence of a 10\,000~K continuum source from an
optically thick accretion disk around the primary to model the
SED. \citet{pet94a,pet94b} increased the estimate of the
temperature of that source to 12\,000~K and suggested that the
cyclic variations in the mass transfer rate could be due to
pulsation of the mass-losing secondary. 
\citet{pet98} obtained a single high-dispersion ORFEUS-SPAS~II far-UV 
spectrum (920 -- 1210\,\ANG) and concluded that it is typical of a B3\,V star 
(the SED including the IUE data can well be fitted with \tef =
17\,000~K and \lgg = 4.0 [cgs]) with \vsin = 120~\ks.

\citet{rich99} made an in-depth study of the types of accretion
structures found in Algol systems. AU~Mon was part of their H$\alpha$
spectroscopic study. They have put AU~Mon in the group with a widely separated
double-peaked disk-like structure.
They have permanent, but variable, accretion disks
similar to those found in cataclysmic variables.
\citet{2005AAS...206.3505M} studied the properties of the accretion structures
in AU~Mon through a multiwavelength spectral study. Double-peaked emission was
detected in the observed H$\alpha$ line confirming the presence of an enduring
accretion disk. The strength of the emission varies with epoch.

In a careful study, \citet{glaz2008} derived new rotational velocities for the
components of 23 close detached and semi-detached binaries. For \au, they
showed how the presence of circumstellar matter can falsify the determination
of the projected rotational velocities.  They attempted to avoid the problem
and obtained $v_{\rm B}$ sin $i$ = 124 \p 4~\ks, and $v_{\rm G}$ sin $i$ =
42.1 \p 2.1~\ks.  They also derived the asynchronicity parameter of the
B-star, $F_B=5.2$.

\subsection{Published light curve solutions} 
To date, a few detailed photometric studies of \au have been published.
Several attempts, all of which are based on Lorenzi's observations prewhitened
for the 411-d variation, were made to derive light curve solutions,
\citet{lorenzi82} himself made the first attempt, creating symmetric normal
points from his light curve and using the Russell-Merrill method.
Independently, and a few months before him, \citet{giu82} solved the light
curve using Wood's model. Finally, another solution, this time based on the
Wilson-Devinney (WD) method \citep{wd71, wilson94}, was published by
\citet{viv98}. These authors concluded that no third light is present in the
system. The main results of these studies are summarized in
Table~\ref{lcsols}.
We note a good agreement between the first
and the third solution, which are based on two independent computer programmes.

\begin{table}
\vspace{0.1cm} \caption[ ]{Published light curve solutions based
on \citeauthor{lorenzi80a}'s \citeyear{lorenzi80a} 
observations.  The symbol $q=M_{\rm
F}/M_{\rm B}$ stands for the mass ratio while the symbols $a$,
$b$, and $c$ denote the relative dimensions of the triaxial
ellipsoids, and the back, side, and pole radii
for the WD model. The following codes are used to identify the
authors: GMM82... \citet{giu82}; L82... \citet{lorenzi82};
VRS98... \citet{viv98}.} \label{lcsols}
\tabcolsep=3pt
\begin{tabular}{lcccrrrrrrrr}
\hline\hline\noalign{\smallskip}
\ \ \ Element & GGM82 & L82 & VRS98 \\
\noalign{\smallskip}\hline\noalign{\smallskip}
$i$ ($^\circ$)&78.4\p0.5&80 assumed &78.74\p0.06 \\
$r_{\rm B}$   &0.115\p0.013&0.18& 0.131\p0.001 \\
$r_{\rm F}$   &  --        &0.18&  -- \\
$a_{\rm F}$   &0.267\p0.056& -- &0.2741\p0.0034 \\
$b_{\rm F}$   &0.242\p0.038& -- &0.2417\p0.003 \\
$c_{\rm F}$   &0.232\p0.031& -- &0.2324\p0.0031\\
$T_{\rm eff}$ (B eq.) (K)&15\,000 assumed &15500 assumed&14500\p1000\\
$T_{\rm eff}$ (B pole) (K)&15\,010& -- &   --   \\
$T_{\rm eff}$ (F eq.) (K)&6\,600\p150&5300&6000\p40 \\
$T_{\rm eff}$ (F pole) (K)&16\,860& -- & -- \\
$q$                  &0.2& -- & 0.1985 \\
$L_{\rm B}$&0.645&0.93 & 0.6590\p0.0051\\
$L_{\rm F}$&0.355&0.07 & 0.3410\\
\noalign{\smallskip}\hline
\end{tabular}
\end{table}


\section{The new photometric data and their analysis}

The new photometric data set consists of a continuous series of 139704
individual \corot photometric observations spanning an interval of 56 days,
i.e., 5 orbital periods\footnote{The \corot data of AU~Mon are public and can
be accessed at \url{http://idoc-corot.ias.u-psud.fr/}}. The spectral domain of
\corot enfolds the range from 370 to 950\,nm, the average time
sampling was 32\,s and roughly $10\%$ of the datapoints were deleted because 
they were extreme outliers. An estimate of the noise level of the
light curve, computed as the average of the periodogram between 80 and 120\,\cd
is $60\,\mu{\rm mag}$.
In addition to the \corot data, we
critically compiled and homogenized all published photoelectric observations
with known dates of observations as well as all available times of minima
known to us and derived 6 new times of primary minima from the \corot light
curve.  Table\,\ref{tab:corotmin} lists the 6 new \corot minima.  The
published visual, photographic, photoelectric and CCD times of minima,
reproduced in Table\,\ref{tab:corotlitmin}, were obtained from the General
Search Gateway of the Variable-Star Section of the Czech Astronomical
Society\footnote{\url{http://var.astro.cz/gsg}} where also references to
original observers can be found.

\begin{table}
\vspace{0.1cm}
\caption{
The 6 new times of minima for AU~Mon derived from the \corot data. 
The O-C values were calculated with our new ephemeris (2) for AU~Mon. 
}\label{tab:corotmin}
\tabcolsep=4pt
\begin{tabular}{lr}
\hline\hline\noalign{\smallskip}
Time (BJD-2454000) & O-C  \\
\noalign{\smallskip}\hline\noalign{\smallskip}

136.67003(1)    &      -0.003        \\
147.77625(1)    &      -0.010        \\
158.89521(1)    &      -0.004        \\
170.00828(2)    &      -0.004        \\
181.12661(1)    &       0.001        \\
192.23762(2)    &      -0.001        \\

\hline
\end{tabular}
\end{table}


\subsection{Time-series analysis and new ephemeris}
\begin{figure}
\centering
\rotatebox{-90}{\resizebox{!}{1.0\hsize}{\includegraphics{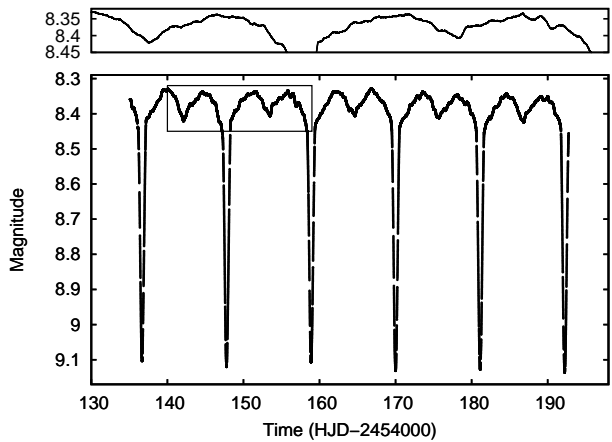}}}
 \caption{The complete \corot light curve of AU~Mon. The upper panel
shows an enlarged segment of the curve with very clear rapid non-orbital
variations. The noise on the data is smaller than the symbol.}\label{corot-t}
\end{figure}

Fig.~\ref{corot-t} shows the complete set of \corot observations vs. HJD. We
converted  \corot fluxes $(F)$ to magnitudes $(m)$ using $m=-2.5\log F + C_0$,
with $C_0$ a calibration constant. We derived $C_0$ through a comparison
between \corot magnitudes and visual magnitudes from literature for all
constant stars in the \corot field of AU~Mon. This gives a value of
$C_0=23.16\pm 0.05$ mag.

First, we verified that Lorenzi's ephemeris (\ref{efe1}) can
reconcile both the compiled published photometry and the \corot data.
Consequently, we used his ephemeris for the initial analyses. In the final
modelling of photometry and radial velocities (RV hereafter)
with the \phoebe programme, described in Sect.~\ref{sec:phoebe},
we derived the following improved orbital ephemeris
\begin{equation}
T_{\rm min\,I} = {\rm HJD}\,2454136.6734(2) +
                  11\fd1130374(1)\times E.\label{efe-new}
\end{equation}

\begin{table*}
\caption[]{Journal of available photometry of \au
with known times of observations.}\label{jouphot}
\begin{flushleft}
\begin{center}
\begin{tabular}{ccrccll}
\hline\noalign{\smallskip}
Station&Time interval& No. of &Passbands&HD of comparison&Source\\
       &(HJD$-$2400000)&obs.  &  &/ check star\\
\noalign{\smallskip}\hline
\hline\noalign{\smallskip}
 1&42790.3--42879.3&   278&$V$  &50109/50346  & \cite{lorenzi80a}\\
 2&43023.6--43843.4&  2143&$V$  &50109/50346  & \cite{lorenzi80a}\\
 3&43877.7--43880.9&   195&$V$  &50109/50346  & \cite{lorenzi80a}\\
 4&44284.4--44287.4&    13&\bv  & all-sky     & \citet{kilk85}\\
 4&44288.3--44289.4&     2&\ubv & all-sky     & \citet{kilk85}\\
 4&44677.3--44678.3&     2&\ubv& all-sky     & \citet{kilk85}\\
 5&45254.9--45322.8&    14&\uvby&50747/50820  & \cite{manfro91}\\
 2&45343.4--45790.4&    82&$V$  &50109/50346  & \cite{lorenzi85}\\
 6&47987.4--49056.1&    81&\hp  & all-sky     & \cite{esa97}\\
 7&54136.0--54192.8&139704& COR & all-sky     & this paper, Corot\\

\noalign{\smallskip}\hline
\end{tabular}\\
\end{center}
\smallskip
{\scriptsize Individual observing stations and photometers distinguished by
the running numbers in column {\sl ``Station":}\\
1... Torino Observatory 1.04-m reflector,  EMI~9502 tube;
2... Torino Observatory 0.60-m reflector,  EMI~6256S tube;
3... Cerro Tololo 0.40-m reflector, EMI~6256 tube;
4... South African Astronomical Observatory (SAAO) 1.0-m \& 0.5-m
     telescopes, EMI~6256 \& 9659 tubes;
5... European Southern Observatory La Silla telescope;
6... Hipparcos Satellite;
7... Corot Satellite.
}
\end{flushleft}
\end{table*}

\begin{figure}
\centering
\rotatebox{-90}{\resizebox{!}{1.0\hsize}{\includegraphics{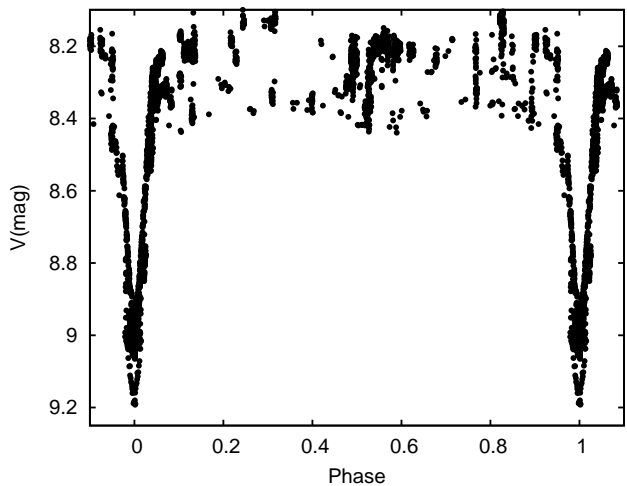}}}
\caption{Original $V$-band light curve 
phased
against the period of 11.1130374 days.  }
\label{newlcv}
\end{figure}

We collected all the available photoelectric observations of AU~Mon and put
them onto a comparable photometric system. The journal of all
observations is shown in Table\,\ref{jouphot}, the data itself can be found in 
Table\,\ref{tab:litphot}. We converted the $uvby$ data 
to Johnson $UBV$ using the transformation derived by \citet{hecboz01}. The
Hipparcos data were transformed to the $V$ magnitude of the Johnson system
using the transformation formula derived by \citet{hpvb}.
Fig.~\ref{newlcv} shows the phased $V$ light curve based on all the published
observations we could collect from the literature, including the 2698
observations provided by \citet{lorenzi80a,lorenzi85}. We can observe a
systematic shift between two extreme states of the maximum light (at the
levels of 8.2 and 8.4 mag respectively) as well as of the primary minima (at
the levels of 9.0 and 9.2 respectively).  This effect is clearly due to the
long-term periodicity reported by \citet{lorenzi80b, lorenzi85}.

\begin{figure}
\centering
\rotatebox{-90}{\resizebox{!}{0.95\hsize}{\includegraphics{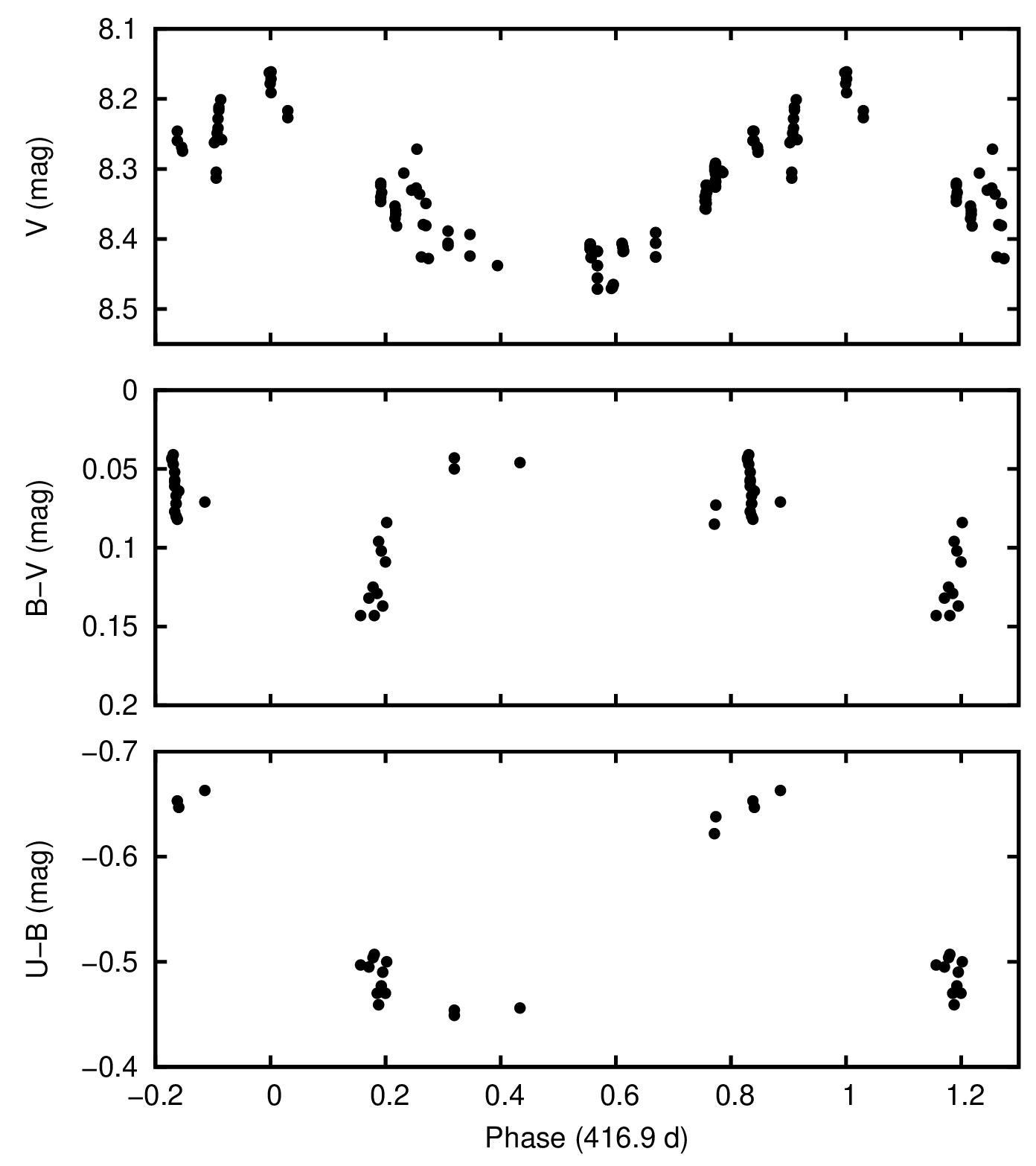}}}
 \caption{The $V$ magnitude of \au outside minima plotted vs. phase
of the 417-d period (top).
The $B-V$ and $U-B$ colour changes along the 417-d cycle of AU~Mon are shown
in the middle and bottom panel (see Sect.\,\ref{sect:long} for details).
}\label{f417}
\end{figure}

Subsequently, using subsets of $V$ data sorted into narrow orbital-phase
bins, as well as all the $V$ data outside of the phase of primary minimum,
we carried out a period search using the PDM \citep{stellingwerf1978} 
and the CLEAN \citep{1987AJ.....93..968R} methods.
Both algorithms yielded the same value of $P_{\rm long}=417\pm8$ days.
Fig.\,\ref{f417} shows the $V$ magnitude outside minima plotted versus phase
of the 417-d period.
Our linear ephemeris for the total brightness of the system reads as follows
\begin{equation}
T_{\rm max. tot. brightness} = {\rm HJD}\,2443105.1(\pm1.4) +
                  416\fd9(\pm8\fd7)\times E.\label{efe-long}
\end{equation}
This also means that the \corot light curve was collected at a phase of total
light minimum (in the phase range from 0.46 to 0.59).

The $V$-band light curves were inspected at the  phases of light minima and
maxima.  The two curves are only shifted for about 0\m2 with respect to each
other (see Fig.\,\ref{newlcv}) but have {\sl the same shape and amplitude}, as
already concluded by \citet{cerruti77}.  This was confirmed by tentative light
curve solutions for both curves, which led to the same values of stellar radii
and orbital inclination.  
This result does not support the
idea promoted by \citet{pet91}, that the observed changes could be related to
apparent changes in the \teff\ and radius of the B-star's false photosphere.
If the 417-d variation were caused by some variable third light, then the
eclipses should become shallower at the 417-d light maxima.  Similarly, a
precession of the orbit would lead to a changing orbital inclination.  Since
none of all these effects is present, we conclude that the long-term variation
must be caused by a periodic attenuation of the total light of the binary by
some variable circumbinary matter. 
The reason why we do not see emission lines
coming from this circumbinary envelope might be that the envelope is cool and
dusty. 
Probably, this huge envelope will also have a small rotational speed. Thus its
emission will be blended with the stronger emission from the circumstellar
matter.


\section{Spectroscopic analysis and radial velocities} 
\subsection{Initial analysis of the spectra\label{sec:initial}} 
New high-dispersion spectroscopic
observations were obtained simultaneously with the \corot data over three consecutive orbital
periods of AU~Mon. They were carried out
in the framework of the \corot ground-based
follow-up programme \citep{2008JPhCS.118a2077U}. The data we used here consist
of 16 FEROS \'echelle spectra
\citep[R$\sim$48000,][]{1997Msngr..89....1K,1999Msngr..95....8K} from the
2.2-m ESO/MPI telescope at LaSilla, Chile, 11 well-exposed \'echelle spectra
secured with the SOPHIE spectrograph
\citep[R$\sim$70000,][]{2008SPIE.7014E..17P} attached to the 1.93-m telescope
of the Haute Provence Observatory, and four spectra taken with FOCES
\citep[R$\sim$65000,][]{1998A&AS..130..381P} at Calar Alto Astronomical
Observatory.  The journal of observations, together with adopted radial
velocities, is given in Table~\ref{jourv}.

All data were subjected to the normal reduction process, which consists of
de-biasing, background subtraction, flat-fielding and wavelength calibration.
All the reduced
spectra were subsequently shifted to the heliocentric frame.
Continuum rectification and cosmic-spike removal were carried out
manually using the SPEFO programme \citep{sef0, spefo}, written by the late
Dr.~Ji\v r\'\i\ Horn 
and until recently being developed by
Mr.~J.~Krpata\footnote{Mr. J. Krpata passed away on February 6, 2009}.

\begin{table}
\caption{Radial velocities measured from the
FEROS (FE), SOPHIE (SO) and FOCES (FO) data. 
The RVs of the G-star were derived via
1-D cross-correlation of suitable parts of the red wavelength region
while the B-star RVs were derived near 4000\,\ANG\ using a 2-D TODCOR-type
cross-correlation. The last three RV columns are the measurements carried out
in SPEFO -- see the text for details. The last column denotes the instrument.
}\label{jourv}
\begin{tabular}{@{}r@{\hspace{1.5mm}}r@{\hspace{1.5mm}}r@{$\pm$}r@{\hspace{1mm}}r@{$\pm$}r@{}r@{\hspace{1mm}}r@{\hspace{1mm}}r@{\hspace{1mm}}r@{}}
\hline\hline\noalign{\smallskip}
 HJD $-$ &S/N&  \multicolumn{2}{c}{RV$_{\rm B}$} & \multicolumn{2}{c}{RV$_{\rm
G}$} &RV$_{\rm He I}$ &RV$_{\rm G}$ &RV$_{{\rm H}\alpha}$ & I. \\
\footnotesize 2454000  & & \multicolumn{2}{c}{\scriptsize(\ks)}  &
\multicolumn{2}{c}{\scriptsize(\ks)}&\scriptsize(\ks)&\scriptsize(\ks)&\scriptsize(\ks)\\
\noalign{\smallskip}\hline\noalign{\smallskip}
76.6493 & 95&   23.3   &  4.2   & -76.74& 3.60& 41.4 & -76.3&   13.0 &FO\\
76.6697 &112&   26.6   &  3.1   & -80.23& 4.06& 42.7 & -78.7&   16.0 &FO\\
78.6665 & 86&   49.8   &  4.4   &-129.55& 3.15& 78.2 &-121.2&   33.1 &FO\\
78.6902 & 86&   59.5   &  6.2   &-128.60& 2.84& 91.3 &-127.0&   43.8 &FO\\
103.7806&117&  -6.553  &  6.939 &  55.94& 1.73&  26.9&  56.4&  -2.7  &FE \\
104.6490&105&  -2.108  & 10.153 & 121.79& 1.04&   0.6& 124.8&  14.2  &FE \\
104.8285& 98&  -7.657  &  8.524 & 132.61& 1.36&  14.0& 137.3&  11.2  &FE \\
105.6557&107& -14.990  &  9.559 & 168.19& 3.30&  -7.5& 175.2&   1.7  &FE \\
105.8796&118& -15.247  &  8.629 & 172.65& 2.37& -20.1& 174.6&   3.4  &FE \\
106.6320& 76& -14.334  &  7.206 & 171.27& 1.57&   1.9& 173.4&   6.1  &FE \\
106.8208&103& -16.287  &  6.911 & 164.85& 1.78& -17.3& 168.3&   9.1  &FE \\
107.7008&124& -10.859  &  4.584 & 125.02& 2.30& -10.3& 126.1& -24.3  &FE \\
108.7038&129&  -2.453  &  0.644 &  42.17& 4.11&   2.0&  44.6&   1.2  &FE \\
108.8141&101&   3.384  &  3.488 &  28.12& 3.99&   3.0&  30.2&  -5.2  &FE \\
108.8739&159&   0.963  &  2.589 &  18.01&11.45&   1.8&  26.1&  -6.7  &FE \\
109.6744& 93&  16.264  &  4.232 & -57.97& 2.87&  23.1& -55.0&  15.1  &FE \\
110.5600&125&  36.924  &  3.668 &-111.92& 2.67&  35.7&-105.7&  27.2  &FE \\
113.3762& 72&  26.498  &  3.610 & -65.29& 1.88&  51.9& -59.4&  74.2  &SO \\
114.3554& 38&  19.044  &  5.689 &  12.74& 1.48&  90.7&  15.2&  61.4  &SO \\
115.3513& 41&  -2.481  &  4.473 &  94.00& 2.63&  20.3&  92.5&  30.7  &SO \\
118.3702& 60& -13.448  &  7.440 & 150.47& 2.79& -26.5& 150.7&  -2.8  &SO \\
119.4596& 53&  -5.039  &  5.894 &  79.44& 3.00& -13.3&  90.6& -23.9  &SO \\
121.3731& 59&  26.867  &  7.542 & -95.42& 3.96&  18.2& -88.1&  41.3  &SO \\
125.7931& 99& -37.327  &  7.089 &  37.10& 2.34&   7.2&  41.0&  31.6  &FE \\
127.6520&103& -12.026  &  9.820 & 161.75& 1.50&   2.2& 164.5&   2.0  &FE \\
128.3489& 58&  -9.204  &  2.343 & 174.38& 1.69&   7.5& 177.8&   --   &SO \\
128.6499&115& -13.868  &  7.290 & 174.36& 2.42&  -2.2& 177.3& -10.0  &FE \\
129.3422& 64& -13.363  &  6.859 & 155.57& 2.06&  -6.6& 158.1& -16.8  &SO \\
130.3320& 62&  -2.870  &  7.891 &  98.79& 1.55&   2.5&  98.4& -28.0  &SO \\
131.3412& 69&   8.946  &  1.713 &  -9.75& 4.66&   5.8&  -2.3&   0.2  &SO \\
132.3469& 57&  27.212  &  6.149 & -87.30& 1.37&  18.5& -85.9&  19.4  &SO \\
\noalign{\smallskip}\hline
\end{tabular}
\end{table}

\begin{figure}
\rotatebox{-90}{\resizebox{!}{1.0\hsize}{\includegraphics{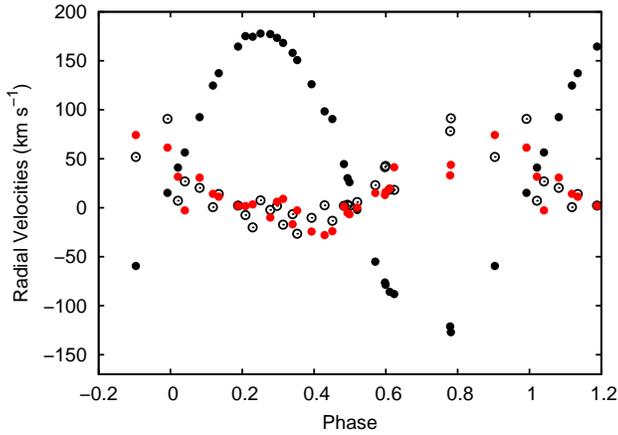}}}
\caption[]{The RV curves of the late-type G-spectrum (black dots),
B-spectrum (open circles) and the \ha emission wings (red dots)
measured in SPEFO. One can see that the B-spectrum RV, based on
the \ion{He}{i}~5876\,\ANG, and that of the \ha emission wings,
define similar RV curves which are both shifted in phase with
respect to the expected RV curve of the G-component.}\label{rvspefo}
\end{figure}

To have some guidance before application of more sophisticated methods,
we first derived the RVs via classical measurements.
Using the programme {\sc SPEFO} 
we carefully rectified all red parts
of  the spectra (between 5500 and 6700\,\ANG) and cleaned them
from cosmics and flaws. 
The RVs of both components were then measured
comparing the direct and flipped line profiles.
For the G-star, we measured \ion{Ca}{i}~6102.723\,\ANG,
\ion{Fe}{i}~6141.730\,\ANG, and \ion{Fe}{i}~6400.000\,\ANG\ which are all
well-defined, unblended, and relatively strong spectral lines.
The r.m.s. errors of the mean RV of these 3 lines ranged from 1 to 4 \ks.
The corresponding orbital RV curve is shown by black circles in
Fig.~\ref{rvspefo}. The only strong and unblended line which seems
to be related to the B-star in the studied red wavelength range
is \ion{He}{i}~5875.732\,\ANG\ (open circles in Fig.~\ref{rvspefo}).
For the \ha line (red dots in Fig.~\ref{rvspefo}), the setting was made
on the steep wings of the double emission and the measurements were
carried out only when a reliable setting was possible.

\begin{table}
\caption[]{Trial circular-orbit solutions for the direct RV measurements
in SPEFO (top). 
{\sc FOTEL} circular-orbit solutions for the RV measurements via
cross-correlation of selected segments of the spectra as described
in Sect.~\ref{rvccf} and tabulated in Table~\ref{jourv} (bottom).
The orbital period was kept fixed at 11\fd1130374. The epoch of
the primary minimum is in HJD-2454000, r.m.s. is the r.m.s. error of 1 observation.}
\label{spefosol}
\tabcolsep=4.7pt
\begin{flushleft}
\begin{tabular}{cccll}
\hline\hline\noalign{\smallskip}
Element  & G-star & \ha emission & \ion{He}{i}\\
         & mean RV&   wings      & absorption \\
\noalign{\smallskip}\hline\noalign{\smallskip}
$T_{\rm min\,I}$&136.668\p0.018&137.94\p0.17&137.46\p0.13\\
$K$ (\ks)       &157.64\p0.16  &32.7\p1.7   &38.6\p1.4   \\
$\gamma$ (\ks)  &22.3\p1.2     &19.4\p2.4   &23.3\p2.4   \\
r.m.s.      (\ks)  &6.19          &12.4        & 12.0       \\
\\[2ex]

\hline\hline\noalign{\smallskip}
Element  & G-star & B-star&\\
         & mean RV& mean RV &\\
\noalign{\smallskip}\hline\noalign{\smallskip}
$T_{\rm min\,I}$&136.684\p0.016&136.972\p0.086&\\
$K$ (\ks)       &158.51\p0.14  &30.41\p0.79   &\\
$\gamma$ (\ks)  &18.8\p1.1     &12.2\p1.1   &\\
r.m.s.      (\ks)  &5.47          &5.98      &    \\
\noalign{\smallskip}\hline
\end{tabular}\\
\end{flushleft}
\end{table}


The RV curve of the G-star is well defined and appears sinusoidal.  To check
on the presence of a possible small eccentricity, we first derived trial
solutions for an elliptical orbit using the programmes SPEL (unpublished,
written by the late Dr. J\'\i\v{r}\'\i\ Horn) and FOTEL \citep{fotel1,fotel2}.
We obtained  $e=0.045\pm0.006$, $\omega=93.\!\!^\circ5\pm7.\!\!^\circ2$, and
$K_{\rm G}=156.71\pm0.92$\,\kms and the test suggested by \citet{lucy71}
indicated that the eccentricity is significant.  However, we do believe that
the orbit {\sl is circular} and the eccentricity {\sl is spurious}, caused by
a small difference between the photocentre of the Roche-lobe filling star and
its centre of gravity and/or by the Rossiter - McLaughlin effect (as detailed
in Sect. 6). It was pointed out by \citet{hec2001,bbin2003} that any
disturbance of the sinusoidal shape of the RV curve which is symmetric with
respect to the systemic velocity {\sl must lead to a formally eccentric orbit}
with a longitude of periastron of either 90$^\circ$ or 270$^\circ$.  This is
exactly what we found. Note also that for these orientations of the binary
orbit, the photometric minima are separated for exactly one half of the
orbital period so that even very accurate photometry cannot provide an
additional test.

Using {\sc FOTEL} \citep{fotel1, fotel2}, we therefore derived a trial
circular-orbit solution for the G star, which is compared to formal solutions
for the RVs of the \ha emission wings and \ion{He}{i}~5876\,\ANG\ absorption
in Table~\ref{spefosol}. The epoch of the primary minimum derived from the
G-star RVs agrees with that of (photometric) ephemeris (\ref{efe-new}). In
contrast to this, the RV curve of the \ion{He}{i}~5876\,\ANG ~absorption and
\ha emission are mutually similar but shifted in phase with respect to the
clean anti-phase orbital motion of the G-component. \citet{popper89} also
suggested such behaviour.  All this is the situation reminiscent of another
W~Ser star with a similar orbital period of 12\fd9, namely $\beta$~Lyr, where
such a behaviour is due to the fact that the absorption and emission lines
originate in bipolar jets emanating from the region of interaction of the gas
streams \citep{hec96}. This may imply that we do not see really photospheric
lines of the hot mass-gaining star of \aue.

\begin{table}
\caption{Basic properties of \au estimated from spectroscopy 
(see the text for details). 
The values of \vsin were derived using the disentangled spectra (see
Sect.\,\ref{sec:tef}), 
the semi-amplitude
${\rm K}_{\rm G}=157-159$~\kms was derived from our spectra and
the orbital inclination was assumed to be in the range from
77$^\circ$ to 83$^\circ$ (see Sect.\,\ref{sec:phoebe}). 
\label{tab:apspec}}
\begin{tabular}{ccc}
\hline\hline\noalign{\smallskip}
  & B-star & G-star \\
  \hline\noalign{\smallskip}
\tef & 15000$\pm$2000~ K & 5750$\pm$250~ K \\[0.5ex]
\lgg & 3.5$\pm$0.3 & -- \\[0.5ex]
M & 6.37$^{+2.18}_{-1.12}$~{M$_\odot$} & 1.17$^{+0.19}_{-0.19}$~{M$_\odot$}  \\[0.5ex]
R & 7.15$^{+5.77}_{-2.92}$~{R$_\odot$} & 9.7$^{+0.6}_{-0.6}$~{R$_\odot$}  \\[0.5ex]
V$_{\rm eq.}^{\rm synch.}$ & 32$^{+26}_{-13}$~\ks & -- \\[0.5ex]
\vsin & 116$\pm$2 \ks & 43.8$\pm$3.5 \ks \\[0.5ex]
$q=M_{\rm G}$/$M_{\rm B}$  & \multicolumn{2}{c}{  0.17$^{+0.03}_{-0.03}$ } \\[0.5ex]
\noalign{\smallskip}\hline
\end{tabular}
\end{table}


\subsection{Mass ratio and secondary mass}

To cope with the above problem, we attempted to obtain some estimate of the
mass ratio independent of the B-star's RV curve.  Since the contact components
of semi-detached binaries usually rotate synchronously with their revolution
\citep[see e.g.][]{mr88}, we made this assumption also for \au and used the
procedure devised by \citet{ander89} as detailed in the Appendix of
\citet{1990A&A...237...91H} to obtain an independent estimate of the mass
ratio ($q$), component masses and the radius (R$_{\rm G}$) of the Roche-lobe
filling secondary.  Its principle lies in the fact that the relative
dimensions of the Roche lobe depend solely on the binary mass ratio $q$ while
the absolute radius of the spin-orbit synchronized secondary is uniquely given
by its equatorial rotational velocity, inclination $i$ of its rotational axis
(assumed to be identical to the inclination of the orbit) and by the
rotational (=orbital) period.  The binary separation is given by the third
Kepler law so that there is only one mass ratio for which the secondary is
just filling the corresponding Roche lobe.  We assumed a circular orbit,
$K_{\rm G}=158\pm1$~\kms and a range of orbital inclinations from 77 to
83$^{\rm o}$ (see Sect.~\ref{sec:phoebe}). We took the uncertainties of the
parameters ${\rm K}_{\rm G}$, \vsini, and $P_{\rm orb.}$ (used in the
procedure) into account to provide the error estimates of the resulting values
of $q$, M$_{\rm G}$, and R$_{\rm G}$.  The results are listed in
Table\,\ref{tab:apspec}.  The mass ratio of AU~Mon equals $q=0.17\pm0.03$. 
Our results agree with \citet[][see
Table\,\ref{lcsols}]{viv98}. \citet{sahade97} found a mass
ratio of $q=0.29$ which is much larger compared to any other study. 
This is due to their usage of the value of $K_B$, which is probably too
optimistic. As Fig.\,\ref{fig:rvsummar} shows, their B-spectrum RVs, from
phase 0.0 to 0.5, are much more negative than any other recorded RVs.
Moreover, our separate circular-orbit solutions for their B- and
G-spectrum RVs give systemic velocities of $1.6\pm2.1$~\ks, and
$5.4\pm3.3$~\ks, differing quite substantially
from the systemic velocity found by us and in all other previous studies.

\subsection{Spectra disentangling and spectral type classification 
\label{sec:tef}} To separate the spectra of both binary
components in an objective way, we used the disentangling procedure developed
by \citet[][see references therein -- Release 2.12.04 of the \korel programme
made available to YF]{korel1,korel2,korel3}.  
The disentangling was applied to different
parts of the spectra. The orbital period was fixed at the value derived by
\citet[][see ephemeris (\ref{efe1}) of the present paper]{lorenzi80b} and a
circular orbit was assumed.

Within the parameter range we investigated, the disentangled spectra were
sufficiently stable to permit to derive the projected rotation velocity and
the effective temperature of both stars. We therefore disentangled the
5500--5700\,\ANG, 6125--6275\,\ANG, and the 4000--4200\,\ANG~wavelength
intervals.  Eleven isolated lines were selected in the secondary's spectrum
and two in the primary spectrum, and we estimated the components' $v\sin i$ by
measuring the position of the first zero of the line profiles Fourier
transform \citep{2002A&A...393..897R}.  The values we found are given in
Table\,\ref{tab:apspec} and are in good agreement with those measured by
\citet{glaz2008}, who tried to avoid the effects linked to the presence of the
companion and of the circumstellar matter. Though our result for the primary
is somewhat smaller, the difference is not significant regarding the scatter
and the small number of lines available in fast rotating early-type stars to
carry out such kind of analysis.  We then compared the component spectra to
synthetic ones, computed for different effective temperatures by means of the
{\sc atlas} LTE model atmospheres \citep{1997A&A...318..841C,
2003IAUS..210P.A20C} and the {\sc synspec} \citep{1995ApJ...439..875H}
programme.  This comparison was done in the red wavelength ranges for the
secondary, while we mainly focused on the bluest range for the primary.  Since
the studied spectral ranges do not show any strong log(g)--dependence, the
surface gravity of the secondary (expected to be evolved) was fixed to \lgg~=
2.5. 

Furthermore, we assumed a microturbulence of 2\,\kms and a solar chemical
composition. Since we used the hydrogen and helium stark-broadened line
profiles for the B-star, our conclusions are in this case not significantly
affected by these assumptions. For the G-star, spectra for different iron
content (i.e. within 0.2\,dex of the solar abundance)  and within 2\,\kms of
the adopted microturbulent velocity were computed and analysed.  The
differences between the models with different iron abundances were found
to be smaller than those obtained by using different effective temperatures
within the estimated error bar of 250\,K. Usually a change in microturbulence
affects more significantly the spectrum. However, the weaker lines which are
present in the different wavelength ranges we have analyzed, and falling in
the linear part of the curve of growth, are not affected by a change in
microturbulence.  These lines were also used to estimate the effective
temperature of the G-star and a good agreement was found between synthetic and
observed spectra.  

\begin{figure}
\rotatebox{-90}{\resizebox{!}{0.95\hsize}{\includegraphics{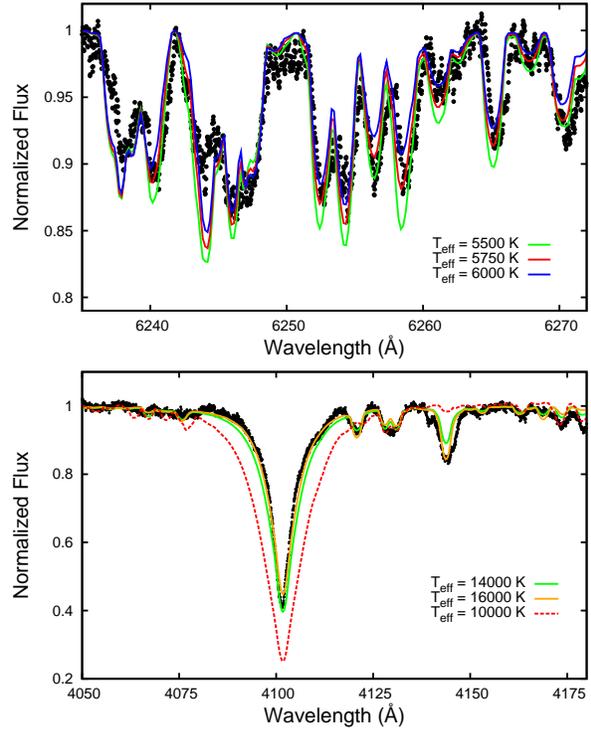}}}
 \caption{ Comparison between synthetic spectra computed for different
effective temperatures and the observed disentangled spectrum (black dots) of
the cool G-component (upper panel) and the B-component (lower panel). The best
fitting temperatures for the G and B-star are 5750\,K and 15000\,K
respectively.
The absorption lines around 4070\,\ANG~are coming from an A0 spectrum (extended
photosphere) and are
best reproduced with a spectrum of $10000\,$K.}
\label{fig:secondloca} 
\end{figure}

\begin{figure}
\begin{center}
\begin{tabular}{c}
\rotatebox{-90}{\resizebox{!}{0.95\hsize}{\includegraphics{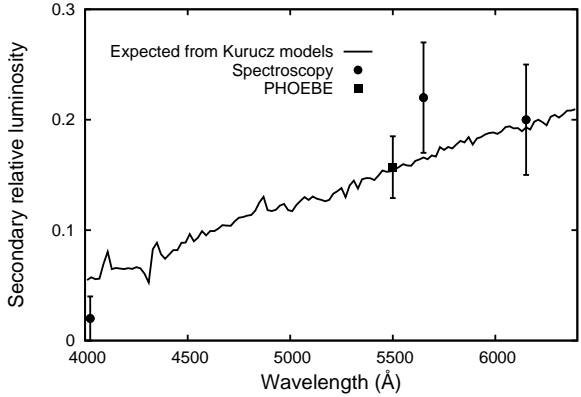}}}
\end{tabular}
\end{center}
 \caption{\label{fig:lumrat} The expected wavelength dependence of
the relative luminosity of the G-star to the total light of the system.
We assumed effective temperatures of 15000\,K and 5750\,K for the B primary
and the G secondary, respectively. The full line is based on Kurucz' models.
The light contribution at 6150\,\ANG\ was fixed at 0.2.
 The luminosity ratio was estimated from detailed comparisons of
the observed and synthetic spectra at several distinct wavelength regions
and these values are shown by black dots with errors. The black square
with an error box represents the luminosity ratio in the $V$ band based
on the \phoebe solution.}
\end{figure}

In Fig.\,\ref{fig:secondloca}, one can see the comparison between the
disentangled and synthetic spectra for a selected wavelength range for the two
components.
The luminosity ratios we estimated in different wavelength domains are plotted
in Fig.~\ref{fig:lumrat}, while the astrophysical parameters we obtained are
listed in Table~\ref{tab:apspec}. As a byproduct of the disentangling
procedure we obtained the RVs for both components (see Fig.\,\ref{fig:rvsummar}).  
For comparison matters, we further assumed
a non-perturbed stellar evolution and a solar-like metallicity. 
To determine the mass, radius and equatorial rotation velocity at orbit
synchronization of the B-star we made use of the evolutionary tracks
calculated by \citet{1992A&AS...96..269S}. 
We interpolated the mass and radius for the effective temperature 
and surface gravity of the B-star throughout 
these evolutionary tracks.
The results of this method can be seen in Table~\ref{tab:apspec}.
The errors on $M_{\rm B}$ and $R_{\rm B}$ are unfortunately 
too large to determine the age
of the star. 
Our result firmly establishes that the cooler star
is a bright G-giant, and not an F-star as was tentatively classified in the
past.

From the lower panel of
Fig.~\ref{fig:secondloca} we see that the agreement between the best
synthetic and observed spectra for the B-star is not
perfect. The B-star spectrum is clearly affected by the
circumstellar matter that acts like an extended photosphere and
produces additional absorption lines generally found in A0 stars
(\tef~$\sim$10000\,K; spectral lines around 4075\,\ANG~are coming from the 
A0 star spectrum).

\subsection{Determination of RVs from whole segments of spectra}\label{rvccf}

Since the motion of the circumstellar matter seems to be
phase-locked with the orbital motion of the B-star, and the
companion is a G-star with many weak but sharp spectral lines, the
radial velocity determination for the G-star is quite
straightforward. To avoid systematic effects due to the presence
of spectral lines belonging to the B-star, we concentrated on the
5500--5850\,\ANG\ wavelength region and carried out
{\sl a 1-D cross-correlation} of the observations with a synthetic
spectrum computed for the stellar spectral classes derived in
Sect.~\ref{sec:tef}. \citep[See][for the correlation maximum
location.]{1995A&AS..111..183D}.
To estimate the accuracy of our measurements,
the cross-correlation was performed separately on 7 subparts of
this spectral range, each sub-region providing one set of radial
velocity measurements. In Table~\ref{jourv} we provide the
mean RVs obtained from those 7 regions and their r.m.s. errors.

\begin{figure}
\rotatebox{-90}{\resizebox{!}{1.0\hsize}{\includegraphics{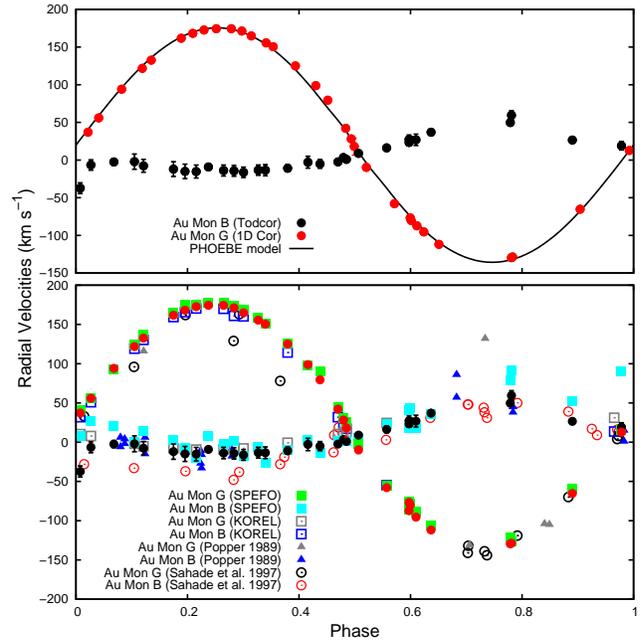}}}
\caption{\label{fig:rvsummar} Radial velocities obtained in this
study by various methods (SPEFO, TODCOR, 1D CC, \korel) compared to published
values. The top panel shows our new and final RVs for sake of clarity, the
bottom panel shows the comparison. The black line shows the final \phoebe 
model for the G-star. The residuals for the G-star are pictured in
Fig.\,\ref{fig:ros}.}
\end{figure}

Due to the orbiting circumstellar matter and the fewer lines
present in the spectrum of early-type stars, the RVs of
the B primary are much more difficult to measure. 
The systematic
study of the whole available wavelength range shows a large
scatter due to: 1. the lack of spectral lines, 2. the presence of
a {\it third spectrum} related to the matter orbiting the primary B-star
(and showing lines generally formed in A0 stars) 
or/and due to bipolar
jets or disk/stream interactions (see Sect.~\ref{sec:initial}). 
The best wavelength
region candidate for the measurement of the primary's radial
velocities is the one that encompassed the \ion{He}{i}~4009 and
4026\,\ANG\ lines where the effects of the (cooler)
{\it third (`A0')} spectrum seem to be absent. Note that the G-star
is more than 3 magnitudes fainter at these wavelengths according
to Fig.~\ref{fig:lumrat}. Still, to reduce any systematic effect due
to the faint G-star spectrum, we used a TODCOR-like procedure
\citep{1994ApJ...420..806Z} using templates that reflect the spectral
types of the components (see Sect.~\ref{sec:tef}). This operation was carried
out several times using different templates for the primary,
with the astrophysical parameters falling within 1-$\sigma$
of the estimated values in order to have an estimate
of the accuracy of the procedure. The results are given
in Table~\ref{jourv} and compared to other published measurements in
Fig.\ref{fig:rvsummar}.

We also derived {\sc FOTEL} circular-orbit solutions for these RVs.
The results are shown in Table~\ref{spefosol}. One can see that the solution for
the G-star agrees reasonably well with that based on the directly measured
RVs (see Table~\ref{spefosol}). The solution for the B-star RVs again shows
a phase shift with respect to the expected instant of the primary minimum
and a significantly lower systemic velocity. The $K$-value for the B-star
agrees within the errors with the $K$-value estimated from our principal
analysis (see Table\,\ref{spefosol}). Nevertheless, because of the phase- and
systematic velocity shift, we will not use the RV curve of the B-star in the
determination of the binary physical properties. 


\section{Modelling of the light and RV data}\label{sec:phoebe}

\begin{table}
\caption{Parameters of AU~Mon together with their formal $3\sigma$
uncertainties 
(standard deviations)  resulting from the
combined \phoebe solution for photometry and RVs. }\label{final}
\begin{flushleft}
\begin{tabular}[sale=0.5]{lcc}
\hline\hline\noalign{\smallskip}
Parameter & Value & Uncertainty\\
\noalign{\smallskip}\hline\noalign{\smallskip}
 HJD$_0$ $(2450000.+) $& $4136.6734$ & $0.0002$\\
 $P_{\rm orb}\,({\rm d})$  & 11.1130374 & 0.0000001\\
 $i\,(\degr)$ & $78.6$    & $0.6$\\
 $q$          & $0.17^a$    & $0.03$\\
 $\gamma$\,(\ks)  & $17.8$    & $0.6$\\
 $A$\,(R$_{\odot}$)  & $42$    & $1$\\
 $e$                 &  $0.01$ & $0.03$\\
 $T_{\rm eff.\,B}\,(K) $  & $14595$    & $900$\\
 $T_{\rm eff.\,G}\,(K) $  & $5750^a$   & --\\
 $\Omega_{\rm B}$   & $7.8$   & $0.9$\\
 $\Omega_{\rm G}$   & $2.12$   & $0.06$\\
 $F_{\rm B}$ & $4.8$   & $0.9$\\
 $[{L_{\rm B} \slash (L_{\rm B} + L_{\rm G})}]_{V}$ & $0.82$ & $0.09$\\
 $[{L_{\rm B} \slash (L_{\rm B} + L_{\rm G})}]_{CoR}$ & $0.84$ & $0.09$\\
 $G_{\rm B}$ & $ 1.0^b$  & --\\
 $G_{\rm G}$ & $ 0.9^b$  & --\\
 $A_{\rm B}$ & $ 1.0^b$  & --\\
 $A_{\rm G}$ & $ 0.5^b$  & --\\
 $x_{\rm B}\,(V)$   & $ 0.458^b$    &0.001 \\
 $x_{\rm B}\,(CoR)$ & $ 0.304^b$    &0.001 \\
 $x_{\rm G}\,(V)$   & $ 0.465^b$    &0.001 \\
 $x_{\rm G}\,(CoR)$ & $ 0.542^b$    &0.001 \\
 $y_{\rm B}\,(V)$   & $ 0.236^b$    &0.001 \\
 $y_{\rm B}\,(CoR)$ & $ 0.195^b$    &0.001 \\
 $y_{\rm G}\,(V)$   & $ 0.243^b$    &0.001 \\
 $y_{\rm G}\,(CoR)$ & $ 0.243^b$    &0.001 \\
 \noalign{\smallskip}\hline\noalign{\smallskip}
 \multicolumn{2}{l}{Roche Radii [in units of orbital separation]} \\
 \noalign{\smallskip}\hline\noalign{\smallskip}
 $r_{B}\,(\rm pole)$ & $0.12$  & $0.02$ \\
 $r_{B}\,(\rm point)$& $0.14$  & $0.02$ \\
 $r_{B}\,(\rm side)$ & $0.14$  & $0.02$ \\
 $r_{B}\,(\rm back)$ & $0.14$  & $0.02$ \\
 $r_{G}\,(\rm pole)$ & $0.22$  & $0.02$ \\
 $r_{G}\,(\rm point)$& $0.31$  & $0.02$ \\
 $r_{G}\,(\rm side)$ & $0.23$  & $0.02$ \\
 $r_{G}\,(\rm back)$ & $0.26$  & $0.02$ \\
 \noalign{\smallskip}\hline\noalign{\smallskip}
 \multicolumn{2}{l}{Absolute dimensions} \\
 \noalign{\smallskip}\hline\noalign{\smallskip}
 \logg$_{\rm B}$    & 3.78      &  0.09\\
 \logg$_{\rm G}$    & 2.50      &  0.03\\
 $M_{\rm bol,B}\,({\rm mag}) $  & $-3.1$   & $0.6$\\
 $M_{\rm bol,G}\,({\rm mag}) $  & $-0.25$   & $0.11$\\
 $M_{\rm B}\,(M_{\odot}) $  & $7.0$  & $0.5$\\
 $M_{\rm G}\,(M_{\odot}) $  & $1.2$  & $0.3$\\
 $R_{\rm B}\,(R_{\odot}) $  & $5.6$   & $0.8$\\
 $R_{\rm G}\,(R_{\odot}) $  & $10.0$  & $0.8$\\
\hline\noalign{\smallskip}
\end{tabular}
\end{flushleft}
{\scriptsize
$^a$ adopted~from~spectroscopy \\
$^b$ assumed}
\end{table}


In order to derive physical properties from the combined light and radial
velocity curves, we used the \phoebe programme, release 031dev \citep[with
phoebe-gui-cairo,][]{prsa05} built on the 2003 WD method \citep{wd71,wil90},
to perform the linearised least-squares analyses with the differential
corrections approach.  We assumed a semi-detached system configuration (using
Mode 5) and no third light nor spots were included.  {\it From previous
considerations, we decided to rely only on the spectroscopic estimation of the
mass ratio and the RVs of the G-star for this modelling. } We also fixed the
effective temperature of the G-star at the value of 5750~K, since this value
was reliably derived from the disentangled spectra (see  Sect.~\ref{sec:tef}). 

The orbital period of 11\fd113037  was initially fixed. For the
bolometric albedos, $A_{\rm B}$ and $A_{\rm G}$, and the gravity darkening
coefficient, $G_{\rm B}$, we used their theoretical values corresponding to
the type of atmosphere and to the spectral types of both stars \citep[see
e.g.][]{2007A&A...471..605V}. For the gravity darkening coefficient $G_{\rm
G}$, we had to adopt a value of 0.9, i.e. much higher than the theoretical
value of 0.32, to obtain a satisfactory fit with the data. This value lies
very close to the value expected for a radiative envelope. 
The limb darkening coefficients $x$ and $y$ of the B- and G-stars were taken
from \citet{2004astro.ph..5087C} for the Johnson $V$ curve while new limb
darkening coefficients specifically computed for the CoRoT passband, which are
now implemented in \phoe (rel. 031dev), were used. The \corot limb darkening
coefficients were derived from a specific extension  of the tables of
\citet{vanhamme93} for the \corot transmission function
\citep[][]{2006ESASP1306.....F},
and computed by Van Hamme (private communication).  The method is described in
\citet{vanhamme93}, it is a convolution of the outgoing intensities with the
\corot transmission curve for different angles.  The errors on $x\,({CoR})$
and $y\,({CoR})$ coming from this method are 0.001.  
The heliocentric Julian
epoch of the primary minimum, HJD$_0$, the effective temperature of the
B-star, $T_{\rm eff.\,B}$, the inclination, $i$, the
dimensionless potential, $\Omega_B$, the fractional luminosity of the
B-star, $L_{\rm B}$, the systematic velocity, $\gamma$,  and the eccentricity,
$e$, were all set as adjustable parameters.  
For the asynchronicity parameter,
$F_B$, we used $4\pm1$ since the projected rotational velocity of
the primary component appears to be $\approx$ 4 times the estimated
synchronous velocity (which amounts to 30 $\pm$ 15 \kms if we adopt the
B-star's radius from the solution presented below). We also checked whether
leaving $F_B$ as a free parameter improved our model. This was not
the case.  

For practical reasons mainly, we constructed 10 different subsets from the
\corot light curve.  Given that the \corot light curve of AU~Mon was largely
oversampled, we have split the continuous time-series into 2 $\times$ 5 full
orbital cycles selecting either even or odd data points in time, thus yielding
2 $\times$ 5 different \corot data subsets.  In this way, no information was
lost and we were able to obtain formal errors based on 10 independent subsets
of the data.  Thus, the 10 \corot data subsets, all the compiled $V$-magnitude
photoelectric observations and the 1D cross-correlation RVs of the G-star from
Table~\ref{jourv} were used in an iterative procedure to search for a
consistent model.  
Of course the \corot data are of much higher quality compared to the $V$-band
light curve which is reflected in the scatter (uncertainties) on the data.

First, we simultaneously modelled each individual \corot data subset with the
past $V$-band light curve.  
In this step we used equal weights both for the \corot and the $V$-band
data.
This rapidly converged towards one possible light curve model obtained from
averaging over the 10 found solutions. The mean parameters (computed with
standard errors) were then adopted as the starting values for the next step in
the modelling.  Next, we simultaneously modelled each individual \corot data
subset and the past $V$-band light curve together with the cross-correlation
RVs of the G-star.  
The weights we assigned in this step to the photometry are now inversely
proportional to the square of the r.m.s. errors for each dataset.  This means
that the \corot data were given a much larger weight $(0.98)$ than the $V$-band
data $(0.02)$ throughout these calculations.  
The reduced
$\chi^2$-values for each modelling approximated the value $\chi^2=1.4$
for the \corot subsets, the fit for the $V$-band light curve had a value of
$\chi^2\approx 2.1$.  Again, an improved light curve and RV model was obtained
from averaging over the 10 found solutions. 

Finally, we improved the orbital period and the epoch of primary minimum, 
HJD$_0$, of \au using \phoebe and
our best-fit model: we simultaneously fitted all the data with the orbital
period and HJD$_0$ set as the only free parameters. This yielded a period of
11\fd1130374(1), which is almost the same as the orbital period determined by
\citet{lorenzi80a} and which we adopted as a fixed parameter in our final
modelling. The final value for HJD$_0$ is HJD$_0=2454136.6734(2)$. 
We checked whether inclusion of third light in the modelling would
improve our solution. This was not the case. 
We also verified 
that, when leaving the mass ratio $q$ as a free parameter,
its value was fully consistent with our spectroscopic estimate of $q$.

\begin{figure}
\rotatebox{-90}{\resizebox{!}{1.0\hsize}{\includegraphics{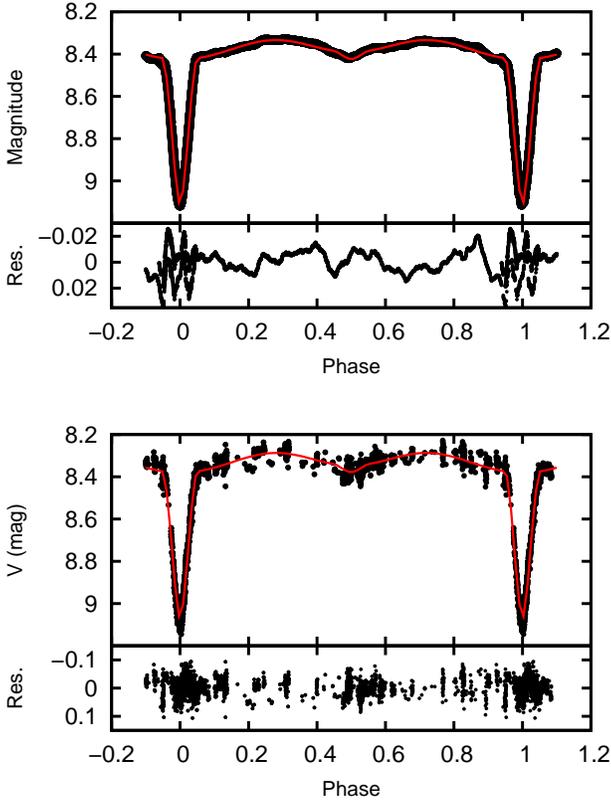}}}
 \caption{The final \phoebe model plotted in red on top of 
one part of the \corot light curve (top) and the $V$-band light curve
(bottom),
after
removal of the long-term period of $P_{\rm long}=417\pm8$ days, phased against
the period of 11.1130374 days.  Below each light curve the residuals of the
model and observations are shown.}\label{phoebe_model_res} 
\end{figure}

\begin{figure} 
\rotatebox{-90}{\resizebox{!}{1.0\hsize}{\includegraphics{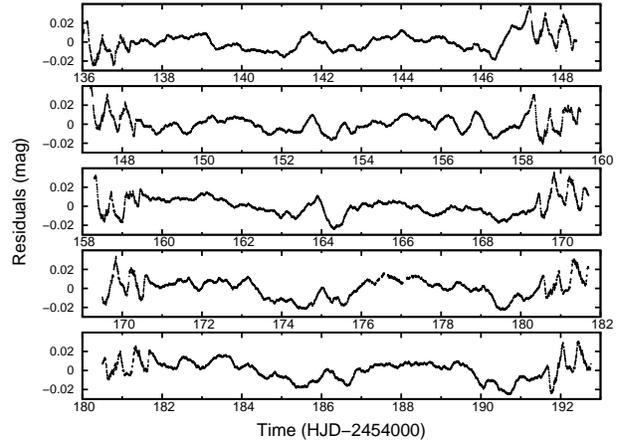}}}
\caption{The \oc\ residuals from the final solution shown for 
the complete \corot light curve. 
}\label{phi_reslc} \end{figure}

\begin{figure} 
\rotatebox{-90}{\resizebox{!}{1.0\hsize}{\includegraphics{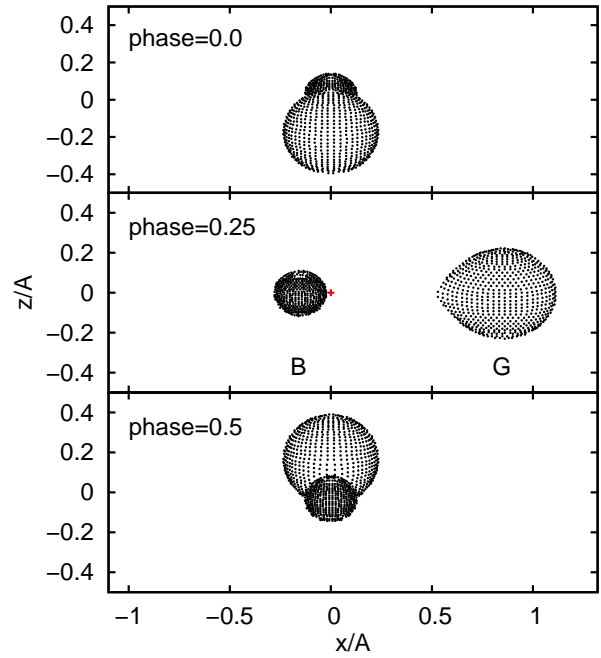}}}
\caption{Representation of the B and G-star of AU~Mon pictured in three 
different phases. The cross in the middle panel denotes the center of mass. 
The system does not show a total eclipse. 
}\label{fig:starplot} \end{figure}

The mean parameters (with estimated uncertainties) and the resulting physical
properties of both components are listed in Table~\ref{final}.  The true
uncertainties will be larger because they depend on unknown systematic
uncertainties and parameter dependencies which were not treated here.  The
observed $V$ and CoRoT light curves and the synthetic model are shown in
Fig.~\ref{phoebe_model_res}.  There is a very good agreement between the CoRoT
observations and the best-fit model: the mean residual value in all the
subsets is of the order of 0.02 mag (Fig.~\ref{phi_reslc}), though somewhat
larger near the phase of primary minimum. 
We note however that the expected effective temperature for a normal
main-sequence B-star with a mass of 7 solar masses would be higher than
15000\,K we obtained, in the order of 20000-21000\,K. The radius of 5.6
solar radii also appears too large, even for the mass of 7 solar masses,
it should be of the order of 4 solar radii. This may indicate that also
from photometry one measures the outer radius of the optically thick
pseudophotosphere, rather than the true radius of the B star. Seen
roughly equator-on, such a pseudophotosphere would also have a lower effective
temperature.
Fig.\,\ref{fig:rvsummar} shows the \phoebe model for the RVs of
the G-star. Fig.\,\ref{fig:starplot} illustrates the configuration of the
binary at three different orbital phases.  

\section{Charaterization of the Rossiter-McLaughlin effect of the G-star}\label{sect:ros}
\begin{figure} 
\rotatebox{-90}{\resizebox{!}{1.0\hsize}{\includegraphics{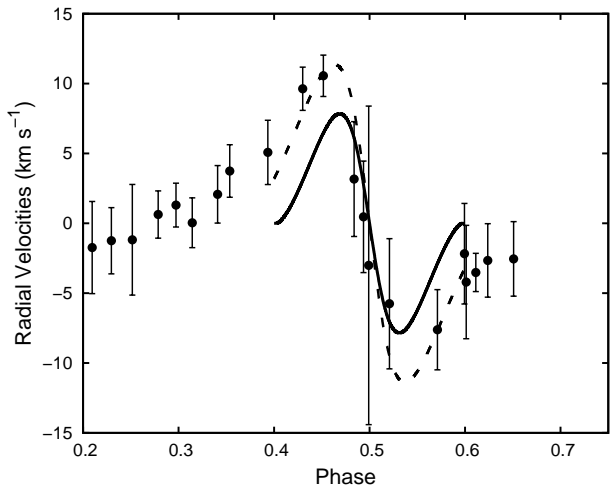}}}
\caption{The orbit subtracted RV residuals (dots) showing the RM effect at
phase 0.5. The solid line is the simulation of the RM effect with the
parameters given in the text, the dashed line shows the RM effect after
an increase of the radius of the G-star. 
}\label{fig:ros} \end{figure}

As already mentioned in Sect.\,\ref{sec:initial}, the RV curve of the G-star
shows a small Rossiter-McLaughlin (RM) effect
\citep{1924ApJ....60...15R,1924ApJ....60...22M} at phase 0.5. This
red-/blue-shifted distortion at the secondary eclipse is due to selective
blocking of the light of the rotating star during an eclipse. When the primary
star covers the blueshifted (redshifted) half of the stellar disk, the
integrated light of the secondary appears redshifted (blueshifted).  Because
of this selective blocking of the stellar surface during the eclipse, a skewed
line profile is created. This change in line profile shape results in a shift
in RV, which in turn results in the redshift-blueshift distortion seen during
the eclipse.  The effect mainly depends on the projected rotation velocity of
the star, the ratio of stellar radii, the orbital inclination, and the limb
darkening.  To analyse this effect we have subtracted the orbital solution
(solid curve in Fig.\,\ref{fig:rvsummar}) from the RV measurements of the
G-star.  The orbit-subtracted RV residuals are shown in Fig.\,\ref{fig:ros}.
We used the analytical description of this effect given in
\citet{2006ApJ...650..408G} to simulate the RM effect. The ratio of the
stellar radii $r_B/r_G$, the inclination and the radius of the G-star relative
to the size of the orbit $r_G$ were taken from our final orbital solution. The
equatorial rotational velocity of the star was set to $44\pm4$\,\ks. The
rotational axis of the G-star is assumed to be perpendicular to the orbital
plane. 

The result of our best fit is seen in Fig.\,\ref{fig:ros}.  With our
current input parameters we cannot fully explain the RM effect.
A possible explanation is that we are primarily not dealing with rotational
effects but with the fact that the photocentre and gravity centre of the
Roche-lobe filling star (G-star) are not identical and coinciding.
\citet{1985ApJ...289..748W} pointed out that this could also cause such
distortions in the RVs during an eclipse.  Another explanation comes from the
fact that we have circumstellar matter together with matter circulating from
the G to the B-component which has not been taken into account in the
calculations.  In order to check this, we have set the relative radius of the
G-star $r_G$ as a free parameter. The result is that we could fit the RM
effect if we increase $r_G$ to $0.44A$ (see Fig.\,\ref{fig:ros}).

\section{Variations on other time scales}
\begin{figure}
\rotatebox{-90}{\resizebox{!}{0.98\hsize}{\includegraphics{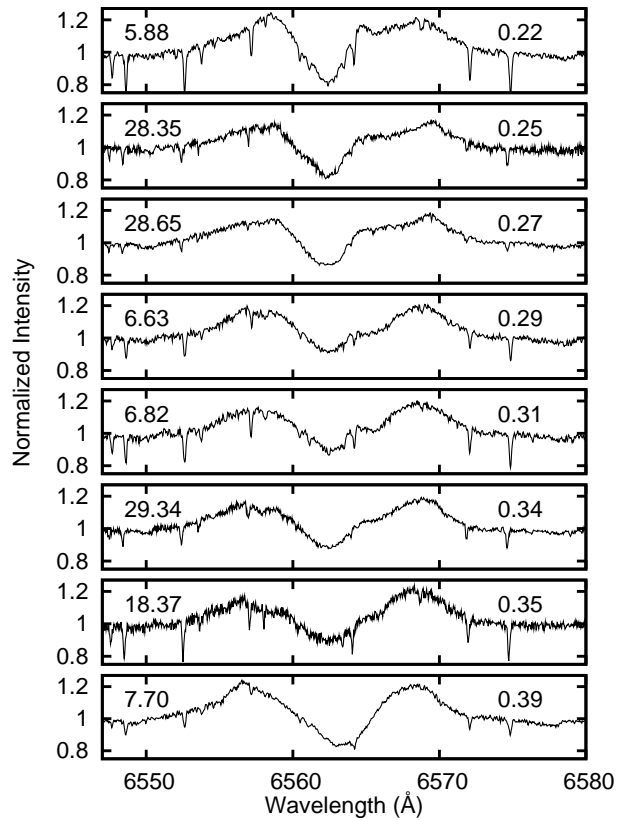}}}
\caption{A sequence of the \ha line profiles showing a complicated velocity
field and cycle-to-cycle variations.  Data from 3 consecutive orbital cycles
can be seen.  In the left corner of each plot we list the time of measurement
(HJD-2454100), in the right corner we give the orbital phase calculated from
ephemeris (\ref{efe-new}). }\label{fig:haprof} \end{figure}

As Fig.~\ref{phi_reslc} clearly shows, a systematic pattern of rapid light
changes which is most pronounced near the phases of primary eclipse was
detected by \corot at every orbital cycle. This is also the reason why the
O-C's calculated from our final ephemeris for these minima are significantly
larger than their accuracy (see Table\,\ref{tab:corotmin}). The  fast changes
near the primary eclipse are reminiscent of what \cite{pavl06} found for W~Cru
(most evident in the U-filter). In accordance with them, we propose that the
variations are due to a non-uniform brightness distribution, probably seated
in the accretion disk.  We also note that the \ha profiles from very similar
orbital phases but from two different orbital cycles differ from each other --
see Fig.\,\ref{fig:haprof}.  For instance, the additional red-shifted
absorption feature is more pronounced in the first cycle than in the last one.


As already mentioned in Sect.\,\ref{sec:spec}, \citet{pet94a,pet94b}
suggested that the cyclic variations in the mass transfer rate could be due to
pulsations of the mass-losing secondary. In order to investigate this we
performed a frequency analysis on the residuals of the light
curve after subtracting our final \phoebe model shown in 
Fig.\,\ref{phoebe_model_res}. 

\begin{figure} \centering
\rotatebox{-90}{\resizebox{!}{1\hsize}{\includegraphics{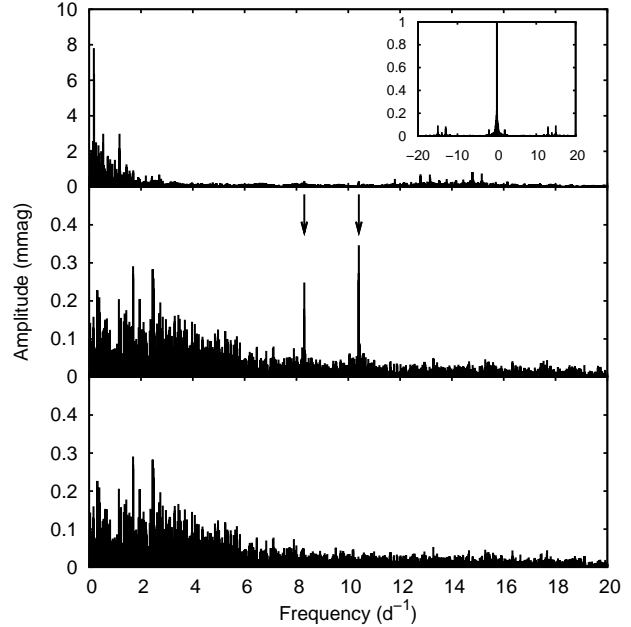}}}
\caption{The periodogram of the residuals out of primary eclipse together
with the spectral window (top+inset). The middle panel shows the periodogram
after prewhitening for 9 low-frequency signals. The bottom panel shows the final
residual periodogram after prewhitening all significant frequency peaks.
}\label{perio_window} \end{figure}

The periodogram of the residual data set is completely contaminated by the
binary orbital frequency and its harmonics. As can be seen in
Figs\,\ref{phoebe_model_res} and \ref{phi_reslc}, the residuals of the \corot
data near the phase of primary minimum show much larger variations than the
residuals outside eclipse.  This periodicity is due to the fact that the
primary minima are not completely well fitted, thus introducing a strong
signal with the orbital frequency of 11.11303 days.  In order to avoid this we
did not consider the residuals during the phase of primary eclipse and
recomputed the Fourier Transform (FT).  The result is shown in
Fig.\,\ref{perio_window}.  The orbital frequency and its harmonics are still
present (from 0 to 3\,\cd) in the top periodogram.  The frequency peaks in the
range of 12 to 17\,\cd are due to the \corot satellite orbital frequency
($f_{\rm orb,CoR}=13.97\,\cd$) and are thus instrumental. We searched for
frequencies using the PDM-method \citep{stellingwerf1978} and prewhitened the
data for the orbital frequency and 8 other low frequency signals ($<0.2\,\cd$)
using a spline fit to the bin means \citep[see e.g.][]{1983A&A...119..279W}.
This resulted in a residual light curve without the binary and satellite
signature. 

\begin{figure} \centering
\rotatebox{0}{\resizebox{!}{1\hsize}{\includegraphics{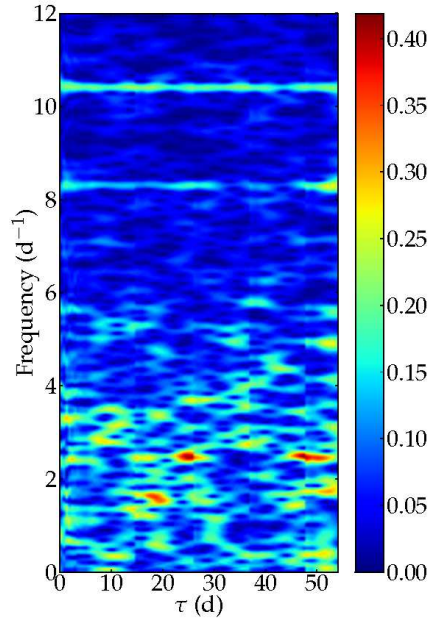}}}
\caption{The Short Time Fourier Transform of the residuals of the \corot light
curve after prewhitening for the orbital frequency. 
The units of the colour scale are in mmag. 
See text for details.
}\label{fig:wavelet} \end{figure}

The residuals from the PDM-prewhitening procedure were analysed using the
Short Time Fourier Transform (STFT) with a Hamming window
\citep{1978ieee...66...51H}.  The result can be seen in
Fig.\,\ref{fig:wavelet}. The STFT was calculated using a window width of 10
days, and evaluated in 100 equidistant time points.  The peaks at 10.4 ($f_1$)
and 8.3\,d$^{-1}$ ($f_2$) are clearly visible as nearly continuous frequency
bands. A short drop in amplitude is noticeable in the 8.3\,d$^{-1}$ frequency
band around day $\tau=38$\,d. The power excess at lower frequency cannot be
attributed to one or several stable frequencies. Instead, the time-dependent
behaviour of low frequencies with short-living amplitudes are visible,
resulting in a smeared out region of power excess in a full Scargle
periodogram (see middle panel of Fig.\,\ref{perio_window}).  It is difficult
to draw firm conclusions about the origin of these features, due to the added
uncertainties from the PDM-prewhitening method in this frequency band.
However the two frequencies in Table\,\ref{TAB:FREQ} are unaffected by our
PDM-prewhitening method, and are definitely present in the Scargle
periodogram. Their characteristics are listed in Table\,\ref{TAB:FREQ}.  
Since these frequencies are not detectable in the spectra, we cannot directly
assign a physical origin to them.  The frequencies $f_1$ and $f_2$ are likely
to originate from the B-star because it produces $\sim82\%$ of the light and
it concerns isolated frequency peaks. Although, theory does not predict such
pulsations for B-stars \citep[see e.g.][]{2007MNRAS.375L..21M}, such
frequencies have also been detected in \corot data of the Be star HD\,49330
\citep{huat2008}. 
This result puts AU~Mon in the neighbourhood of the sample of semi-detached
Algol-type eclipsing binaries with an oscillating mass accreting component
\citep[oEA stars, with RZ~Cas as the best studied
object][]{2007ASPC..370..194M}.  AU~Mon can not be fully classified as an oEA
star, because oEA stars are at the end of the mass transfer regime which is
still not the case for AU~Mon.  

On the other hand, theory predicts that the G-star should produce solar like
oscillations \citep[see e.g.][]{2002MNRAS.336L..65H,2007A&A...463..297S}.
Since we know the mass, radius and $T_{\rm eff}$ of the star, we predict
$\nu_{\rm max}=M_G/R_G^2\cdot(T_{{\rm
eff},G}/5777)^{-0.5}\cdot3050\approx34\,\mu {\rm Hz}~(=2.94\,\cd)$ to be the
frequency at which to expect such oscillations \citep{1995A&A...293...87K}.
This value is in agreement with the location of the power excess in the
Scargle periodogram of AU~Mon (middle panel of Fig.\,\ref{perio_window}). The
amplitude of the power excess is not in disagreement with the amplitudes of
the discovered solar-like oscillations in the \corot data of red giants
\citep{2009Natur.459..398D} taking into account the flux contribution of the
G-star.  Due to a lack of mode identification and uncertainty on the origin of
the oscillations, we are unable to exploit the detected frequencies
seismically without additional spectroscopic information about them.

\begin{table}
\caption{Frequencies found in the residuals
together with their S/N ratio (we refer to the text
for explanation). The error estimate \citep{1999DSSN...13...28M}
for the frequencies is $\pm
0.0002\,\cd$. The error on the
amplitude is
$0.01$ mmag.
}
\label{TAB:FREQ}
  \begin{tabular}{lcccc}
\hline \hline
ID &\multicolumn{2}{c}{Frequency} & Amplitude & S/N \\
& [\cd] & [\mHz] & [mmag] &\\
\hline
$f_1 $ &  10.4081  & 120.46  & 0.34   & 14.8  \\
$f_2 $ &  8.3033   & 96.10  & 0.25   & 9.2  \\
\hline

\end{tabular}
\end{table}

\section{Understanding the long-term light variations}\label{sect:long}
Other cases than \au exist in which the discovery of long-term periodic or at
least cyclic brightness variations (reminiscent of \au) have been claimed,
namely those cases where systematic photometry in a calibrated photometric
system has been carried out or where interacting binaries with Balmer emission
lines and hot mass-gaining components are discussed.
Probably the first reports of such a behaviour are RX~Cas \citep{kalv79},
TV~Cas = HD~1486 \citep{walter79} and V505~Sgr =BD$-14^\circ5578$
\citep{walter81}.  However, the latter two systems exhibit only very faint
single-peaked \ha emission and their long-term variations are probably related
to the presence of distant third components in these systems \citep{rich99,
ves2001}.

\begin{table}
\caption[]{Galactic hot emission-line binaries with cyclic long-term
brightness changes. The tabulated mass ratio is always the
ratio of the mass of the lighter (usually Roche-lobe filling)
component to that of the B-star.}\label{twoper}
\tabcolsep=3.2pt
\begin{flushleft}
\begin{tabular}{rrcccll}
\hline\hline\noalign{\smallskip}
Name   & BD & $P_{\rm orb.}$ & $P_{\rm long}$ & Mass ratio & Ref.\\
\noalign{\smallskip}\hline\noalign{\smallskip}
 RX Cas    &$+67^\circ244 $&32\fd312&516\fd06  & 0.30\p0.05 & 1, 2, 3, 4\\
 AU Mon    &$-01^\circ1449$&11\fd113&417\fd0   & 0.17\p0.03& 5 \\
 CX Dra    &$+52^\circ2280$&6\fd696 &$130^{\rm d}$--$180^{\rm d}$& 0.23       & 6, 7, 8 \\
$\beta$~Lyr&$+33^\circ3223$&12\fd94 &282\fd37  & 0.225      & 8, 9, 10 \\
V360 Lac   &$+41^\circ4623$&10\fd085&322\fd24  & 0.163      & 11, 12 \\
\noalign{\smallskip}\hline
\end{tabular}\\
\smallskip
{\scriptsize References to original studies identified by the running numbers
in column ``Ref.":\\
1... \citet{kalv79};
2... \citet{kriz80};
3... \citet{ander89};
4... \citet{pust2007};
5... this study;
6... \citet{zarf10};
7... \citet{horn92};
8... \citet{rich2000};
9... \citet{hec96};
10... \citet{hec2002}
11... \cite{zarf17};
12... \cite{zarf24}
}
\end{flushleft}
\end{table}

In Table\,\ref{twoper}, we provide basic information on galactic emission-line
systems with cyclic long-term variability known to us.
During the past five years, similar systems were also discovered and studied
rather intensively in the Magellanic Clouds \citep{men2003, men2005a,
men2005b, men2006, men2008}.
In spite of all effort, the true cause of the long-term cyclic changes remains
unexplained, although several different interpretations were put forward:
\citet{kalv79} interpreted the 516-d periodicity of RX~Cas as pulsation of the
Roche-lobe filling star.  \citet{hec96} considered the possibility that the
282-d cycle of $\beta$~Lyr is a beat period between the orbital period and
rapid changes with a cycle of 4\fd7 which they detected in spectroscopy. They
also pointed out some similarity with the 164-d cycle of V1343~Aql = SS~433, a
massive 13\fd08 X-ray binary with bipolar jets. \citet{wils99} also studied
$\beta$~Lyr. They concluded that neither apsidal advance nor precession can
account for the 282-d light variation. They were unable to exclude pulsations
of the disk as the cause. As already mentioned, \cite{pet91} concluded that
the optical brightness variations of \au are due to a 1200~K variation in the
photospheric effective temperature of the B primary.

\citet{men2003} reported the discovery of eclipsing binaries with long-term
periodic brightness changes from OGLE photometry of the Magellanic Clouds.
\citet{men2003} pointed out that the light maxima of
the long cycles are always accompanied by a mild reddening of the objects in
question. \citet{men2005a} concluded that the long brightness changes are
often cyclic rather than strictly periodic ones.  
Subsequently, \citet{men2008}
concluded that the long-term variations must be due to variations in {\it
circumbinary} matter as we found for \aue. They suggested that the system
experiences supercycles of mass outflow which lead to replenishment of the
circumbinary envelope.

AU~Mon may become a key object to study the true nature of the long-term
brightness changes since it is bright and the amplitude of the 417-d period is
large.  In spite of our effort to collect and homogenize existing photometry,
available material on colour variations is, unfortunately, very scarce.  In
Fig.\,\ref{f417}  we compare the $V$ magnitude and $B-V$ and $U-B$ colour
changes along the 417-d cycle for all observations we were able to transform
into comparable $UBV$ magnitudes.  Unfortunately, there is no colour
information near phases of the brightness maximum.  A plot of both colour
indices vs. time shows that there could also be colour variations on
longer time scales, than those related to the 417-d cycle. If this is a more
general pattern, this could perhaps also explain somewhat contradictory
reports on the colour behaviour of various objects studied by Mennickent and
his collaborators.  Clearly, future systematic calibrated multicolour
photometry of AU~Mon over the whole 417-d cycle is very desirable.

We tentatively suggest that the long-term brightness
changes of \au  must be associated with some circumbinary
matter and it is plausible to assume that the bulk  of such
material is associated with putative bipolar jets for which we find
indirect support from the shift of the RV curves of some spectral
lines associated with the B-star. 
\citet{2000ASPC..214..697B} showed via hydro calculations for $\beta$ Lyr that
when the encircling stream hits the denser primary stream from the mass-losing
star, it gets bended and goes out of the orbital plane resulting in jet-like
structures. 
\citet{pet2007} reported probable detections of bipolar jets of very
hot plasma, perpendicular to the orbital plane, for three other hot
interacting binaries: TT~Hya, V356~Sgr and RY~Per. However,
\citet{2007ApJ...656.1075M} studied circumstellar matter of 
TT~Hya and argued against jets.
The variations themselves could have two possible causes: either
cyclic changes in the mass outflow from the binary as suggested by
\cite{men2008} or precession of the binary orbit which would
change the attenuation of the binary due to changing projection
effects of jets. This latter idea seems improbable, however, in the
light of our finding that the light curve from the maxima and minima
of the 417-d period lead to the same binary elements including the
orbital inclination. It is clear that continuing systematic observations
of \aue, including spectro-interferometry and polarimetry, could help
to understand the nature of the remarkable changes of these interesting
objects.

\section{Summary}
Our analyses of very accurate CoRoT space photometry, past Johnson $V$
photoelectric photometry and high-resolution \'echelle spectra led to the
determination of improved fundamental stellar properties of both components of
the massive and interacting system AU~Mon. We derived new and accurate ephemerides for
both the orbital motion (with a period of 11\fd1130374) and the long-term,
overall brightness variation (with a period of 416\fd9). It is shown that this
long-term variation must be due to attenuation of the total light by some
variable circumbinary material.  We derived the binary mass ratio $M_{\rm
G}/M_{\rm B}$ = 0.17\p0.03.  Using this value of the  mass ratio as well as
the radial velocities of the G-star, we obtained a consistent and coherent
light curve solution and a new estimate of the stellar masses, radii,
luminosities and the effective temperatures.  

We must point out that our final model does not include the gas stream and
accreting matter on the B-star.  It would be interesting to consider such
complications in any future modelling of the binary.  We also report the
discovery of rapid and periodic light changes visible in the high-quality
residual CoRoT light curves.  The rapid light changes visible in the residuals
near primary minima  repeat at every orbital period. 
They are probably due to a non-uniform brightness distribution, seated in
the accretion disk.
Outside the
primary minima of the \corot light curve we detect two frequencies, they
are in the expected frequency domain of B-stars. 
Complementary interferometric and polarimetric
observations will be needed to even better understand the geometry and the
nature of the circumbinary matter in \aue.

\section*{acknowledgements}
The authors acknowledge critical remarks and useful suggestions by Dr.
Johannes Andersen on an earlier version of the paper.
YF thanks P. Hadrava for making  his \korel computer code (Release 2.12.04)
available to him. We also thank P. Hadrava for the use of his FOTEL computer
code. This research has received funding from the European Research Council
under the European Community's Seventh Framework Programme
(FP7/2007--2013)/ERC grant agreement n$^\circ$227224 (PROSPERITY), as well as
from the Research Council of K.U.Leuven grant agreement GOA/2008/04.  MB is
Postdoctoral Fellow of the Fund for Scientific Research, Flanders.  The
research of PH was supported by the grants 205/06/0304 and 205/08/H005 of the
Czech Science Foundation and also from the Research Programme MSM0021620860 {\sl
Physical study of objects and processes in the solar system and in
astrophysics} of the Ministry of Education of the Czech Republic.  The FEROS
data are being obtained as part of the ESO Large Programme: LP178.D-0361 (PI:
Poretti). This work was supported by the Italian ESS project, contract
ASI/INAF I/015/07/0, WP\,03170. KU acknowledges financial support from a {\sl
European Community Marie Curie Intra-European Fellowship}, contract number
MEIF-CT-2006-024476.  PJA acknowledges financial support from a ``Ram\'on y
Cajal'' contract of the Spanish Ministry of Education and Science.  We
acknowledge the use of the electronic data base {\sc SIMBAD} operated by the
CDS, Strasbourg, France and the electronic bibliography maintained by the
NASA/ADS system.  Finally, we thank Andrea Miglio for valuable discussions on
stellar oscillations.


  \bibliographystyle{mn}
  \bibliography{aumon,aumono,new}

  \bsp


\begin{appendix}
\section{Times of minima}\label{ap1}



\begin{table}
\vspace{0.1cm}
\caption{
Visual, photographic, photoelectric and CCD times of minima.  The times were
adopted form the General Search Gateway of the Variable-Star Section of the
Czech Astronomical Society ({\sl http://var.astro.cz/gsg}).  The O-C values
were calculated with our new ephemeris (\ref{efe-new}) for AU Mon.  The third
column lists the type of observation. vis: visual; pg: photographic; pe:
photo-electric.
}\label{tab:corotlitmin}
\tabcolsep=4pt
\begin{tabular}{lrc@{\hspace{8mm}}lrc}
\hline\hline\noalign{\smallskip}
Time  & O-C & Type&Time  & O-C & Type \\
(HJD-2400000) &  & & (HJD-2400000) &  &  \\
\noalign{\smallskip}\hline\noalign{\smallskip}

26743.160    &       0.124    &  vis  &37378.364    &       0.151    &  pg   \\
26743.390    &       0.354    &  pg   &37400.351    &      -0.088    &  pg   \\
26765.320    &       0.058    &  pg   &37578.633    &       0.385    &  pg   \\
26976.400    &      -0.010    &  vis  &37667.505    &       0.353    &  pg   \\
26987.530    &       0.007    &  vis  &38322.601    &      -0.220    &  pg   \\
27076.430    &       0.003    &  vis  &38378.568    &       0.182    &  pg   \\
27120.890    &       0.011    &  vis  &38411.508    &      -0.217    &  pg   \\
27487.583    &      -0.027    &  vis  &38467.353    &       0.062    &  pg   \\
27843.110    &      -0.117    &  vis  &38856.363    &       0.116    &  pg   \\
27887.631    &      -0.048    &  vis  &39200.388    &      -0.363    &  pg   \\
27932.230    &       0.099    &  pg   &39945.411    &       0.086    &  pg   \\
28209.912    &      -0.045    &  vis  &40478.609    &      -0.141    &  pg    \\ 
28498.919    &       0.023    &  vis  &40656.362    &      -0.197    &  pg    \\ 
28543.377    &       0.029    &  vis  &41056.332    &      -0.296    &  pg    \\ 
28921.155    &      -0.037    &  vis  &41334.391    &      -0.063    &  pg    \\ 
28954.480    &      -0.051    &  vis  &41356.438    &      -0.242    &  pg    \\ 
28965.656    &       0.012    &  vis  &41367.466    &      -0.327    &  pg    \\ 
30021.380    &      -0.002    &  pg   &41390.355    &       0.335    &  pg    \\    
30810.400    &      -0.008    &  pg   &42090.373    &       0.232    &  pg    \\    
31910.610    &       0.011    &  pg   &42790.2350    &      -0.027    &  pe    \\   
32888.560    &       0.014    &  pg   &42801.3602    &      -0.015    &  pe    \\   
33266.400    &       0.011    &  pg   &42868.1210    &       0.068    &  pe    \\   
33533.160    &       0.058    &  pg   &42879.2360    &       0.069    &  pe    \\   
34444.380    &       0.009    &  pg   &43123.653    &      -0.000    &  UBV    \\   
34811.100    &      -0.001    &  pg   &43190.332    &       0.001    &  UBV    \\   
35044.500    &       0.025    &  pg   &43201.449    &       0.005    &  UBV    \\   
35900.386    &       0.207    &  pg   &48502.220    &      -0.143    &  V    \\     
37367.333    &       0.233    &  pg   &51513.5290    &      -0.468    &  ccd    \\  
\hline
\end{tabular}
\end{table}

%
%
%
%
%


\section{Available photoelectric data}

\begin{table}
\vspace{0.1cm}
\caption{
Excerpt of the available photoelectric observations of AU~Mon.
The full Table is available as Supporting Information.
}\label{tab:litphot}
\tabcolsep=4pt
\begin{tabular}{lr}
\hline\hline\noalign{\smallskip}
\multicolumn{2}{l}{Data from \citet{lorenzi85}}\\
\hline\noalign{\smallskip}
Time  & $V$ \\
(HJD-2400000) & (mag)  \\
\noalign{\smallskip}\hline\noalign{\smallskip}
45343.4114 &  8.380  \\
45346.3339 &  9.137  \\
45346.4262 &  9.048  \\
45351.4059 &  8.415  \\
45351.4202 &  8.418  \\
45353.3960 &  8.391  \\
45354.4253 &  8.387  \\
45355.3607 &  8.385  \\
45356.3518 &  8.448  \\
45357.2635 &  9.118  \\
45357.2823 &  9.151  \\
45357.3028 &  9.165  \\
45357.3205 &  9.181  \\
45357.3349 &  9.180  \\
45357.3471 &  9.187  \\
45357.3600 &  9.188  \\
45357.3739 &  9.185  \\
45357.3927 &  9.192  \\
45357.4045 &  9.180  \\
45357.4161 &  9.169  \\
45357.4287 &  9.165  \\
45357.4403 &  9.155  \\
45357.4565 &  9.147  \\

\hline
\end{tabular}
\end{table}

\section*{Supporting Information}
Additional Supporting Information may be found in the online version of this
article:
\\
{\bf Table\,\ref{tab:litphot}.} Available photoelectric observations of
AU~Mon.

~~

\end{appendix}

\label{lastpage}
\end{document}